\documentstyle[11pt,epsf]{article}

\def\beq{\begin{eqnarray}}    
\def\eeq{\end{eqnarray}}      

\def\Tr{\,\mbox{Tr}\,}                  
\def\pa{\partial}                       
\def\al{\alpha}
\def\be{\beta}

\def\ga{\gamma}
\def\de{\delta}
\def\vp{\varepsilon}
\def\ep{\epsilon}
\def\ze{\zeta}

\def\la{\lambda}
\def\na{\nabla}

\def\si{\sigma}

\def\ph{\varphi}

\def\Ga{\Gamma}
\def\De{\Delta}

\newpage
\begin{document}
\begin{titlepage}

\begin{flushright}
\end{flushright}


\begin{center}

\vspace{1cm}
{\LARGE  \sl Torsion action and its possible observables}
\vskip 2mm

\vskip 7mm
\setcounter{page}1
\renewcommand{\thefootnote}{\arabic{footnote}}
\setcounter{footnote}0

{\bf A.S. Belyaev} $^{\mbox{a,b,}}$\footnote{e-mail:
belyaev@ift.unesp.br},
$\,\,\,${\bf I.L. Shapiro} $^{\mbox{c,d,}}$\footnote{e-mail:
shapiro@ibitipoca.fisica.ufjf.br}

\vspace{7mm}

$^{\mbox{a}}$ {\it Instituto de F\'\i sica Te\' orica, Universidade
Estadual Paulista,\\
Rua Pamplona 145, 01405-900 - S\~ao Paolo, S.P., Brazil}

\vspace{3mm}

$^{\mbox{b}}$ {\it Skobeltsyn Institute of Nuclear Physics,
Moscow State University, \\ 119 899, Moscow, Russian Federation}

\vspace{3mm}

{\it $^{\mbox{c}}\,\,$
Departamento de F\'\i sica -- Instituto Ciencias Exatas, 
Universidade Federal de Juiz de Fora,\\
36036-330, Juiz de Fora - MG - Brazil }

\vspace{3mm}

{\it $^{\mbox{d}}\,\,$ 
Tomsk State Pedagogical University, 634041, Tomsk, Russia }

\vspace{7mm}
\end{center}

\begin{abstract}
The methods of effective field theory are used to explore the
theoretical and phenomenological aspects of the torsion field.
Spinor action coupled to electromagnetic field and torsion
possesses an additional softly broken gauge symmetry. This
symmetry enables one to derive the unique form of the torsion action
compatible with unitarity and renormalizability. It turns out 
that the antisymmetric torsion field is equivalent to an 
massive axial vector field.
The introduction of scalars leads to serious problems which
are revealed after the calculation of the leading two-loop 
divergences.
Thus the phenomenological aspects of torsion may be studied only
for the fermion-torsion systems. In this part of the paper we obtain
an upper bounds for the torsion parameters using present experimental
data on forward-backward Z-pole asymmetries, data on
the experimental limits on four-fermion contact interaction
(LEP, HERA, SLAC, SLD, CCFR) and also TEVATRON limits on the cross
section of new gauge boson, which could be produced as a resonance
at high energy $p\bar{p}$ collisions. The present experimental data
enable one to put the limits on torsion parameters for the various range 
of the
torsion
mass. We emphasize that for the torsion mass of the 
order of the Planck mass no any independent theory for torsion is 
possible, and one must directly use string theory.

\end{abstract}

 \vskip 0.5cm
\end{titlepage}

\vskip 6mm
\noindent
{\large\bf Introduction}
\vskip 2mm

Despite the impressive success of the Standard Model (SM) there are many
reasons to suspect
that it doesn't cover nor all of the existing fields neither all the
interactions. The main reason is that the SM unifies strong, weak and
electromagnetic interactions while the quantization of gravity remains
beyond its scope. It is commonly accepted that the consistent quantum
gravity does not exist and that, instead of quantizing gravity itself
one has to start with the fundamental theory of different nature which
should produce gravity as a low-energy effective theory. The theory of
string is a nice example of such a theory: it is consistent as a quantum
theory and it induces the gravitational
interaction in the low-energy limit (see, for example, \cite{GSW}).
However, along with a metric, the string
theory also predicts other constituents of gravity: in particular
those are: a scalar field -- dilaton and antisymmetric tensor field
of the third rank which is usually associated with torsion. The studies
of gravity with torsion have a long history (see, for example,
\cite{hehl}) but now the investigation of the possible effects and
manifestations of torsion becomes more actual since this can be a
possible way to understand the low-energy effects of string physics.

The main theoretical advantage of gravity with torsion is that it
links the spin of the matter fields with the space-time geometry.
This feature was in a heart of the most of the works about the
possible effects of torsion. The classical and quantum
theory of fields and particles in an external gravitational field with
torsion has been studied in a papers 
\cite{dat,
aud,prec,hehl,kernel,creation,bush1,bush2,babush,
hammond,doma1,lamme,rytor}
(the reader is referred to these publications and to the review 
\cite{hehl-review} for further 
references and also
to the book \cite{book} for the introduction to quantum field theory in
an external gravitational field with torsion). The quantum field 
theory in
an external torsion field predicts many physical effects which we are
not going to review here.
It is important for us that the renormalizability
of quantum theory in an external torsion field requires the nonminimal
interaction of an external torsion with a spin-$\frac12$ field and also
with a scalar field which doesn't interact with torsion in a minimal way
\cite{bush1}. Thus the interaction of torsion with the spinor field is
characterized by the special coupling constant similar to the electron
charge. Until now we do not have any positive data about the
value of this parameter or about the order of magnitude that the 
external
torsion may have. Moreover without the full theory including the
dynamical equations for the torsion itself one can not understand how 
the
external torsion field can be generated in the laboratory.
In this situation one promising possibility is to look for an
external torsion field which can exist in our part of the Universe.
Despite the existing upper bounds for such a field are quite rigid
\cite{lamme} it can be taken responsible \cite{doma1} for the recently
observed anisotropy of the polarization of light coming from some cosmic
objects \cite{rano}.

Independent on the development of the classical and quantum
field theory in an external torsion field it is important to
 establish the form of the action for the torsion itself and to
study the possible experimental effects of dynamical torsion.
There are, indeed,
several known possibilities to approach this problem. In the famous
work of Kibble \cite{kibble} the action of gravity with torsion
has been derived as an action of the compensating fields for the
general coordinate transformation. Unfortunately this action doesn't
lead to consistent quantum theory since the last is nor unitary neither
renormalizable, despite it contains higher derivatives.
In the papers \cite{nev,sene} many versions of the action of gravity
with torsion had been constructed using the condition of unitarity.
All those actions are of the second order in curvature and are similar
to the Lovelace action which includes the Gauss-Bonnet topological
term\footnote{Later on similar actions have been used by Zweibach et al
\cite{zwei} for constructing string effective actions which do not
have problems with unitarity. In the last of the papers \cite{zwei}
the string effective actions with torsion were also discussed.}.
Unfortunately those actions are very constrained and the
corresponding theory has the renormalizability properties
even worst than the action of Kibble.

The problem of renormalizability is casted in another form if we
consider it in the content of effective field theory 
\cite{weinberg,dogoho}.
In the framework of this approach one has to start with the action
which includes all possible terms satisfying the symmetries of the
theory. Usually such an action contains higher derivatives at least in
a vertices. However as far as one is interested in the lower
energy effects, those high derivative vertices are suppressed by the
huge massive parameter which should be introduced for this purpose.
Then those vertices and their renormalization is not visible and
effectively at low energies one meets renormalizable and unitary theory.
The gauge invariance of all the divergences is guaranteed by the
corresponding theorems \cite{volatyu} and thus this scheme may be
applied to the gauge theories including even gravity \cite{don}.
Within this approach it is important that the lower-derivative
counterterms have the same form as a terms included into
the action. This condition, together with the symmetries and the
requirement of unitarity, may help to construct the effective field
theories for the new interactions which are not yet observed, but
anticipated from the development of the fundamental theories like
string.

If one starts to formulate the dynamical theory for torsion in this
framework, the sequence of steps is quite definite. First one has
to establish the field content of the dynamical torsion theory and
the form of interactions between torsion and other fields. Then it
is necessary to take into account the symmetries of this interactions
and formulate the action in such a way that the resulting theory
is unitary and renormalizable as an effective field theory.
Indeed there is no guarantee that all these requirements are consistent
with each other, but the inconsistency may only indicate that some
symmetries are lost or that the consistent theory with the given
particle content is impossible.

In this paper \footnote{An early report including part of our studies
has been published in \cite{betor}.} we follow the above scheme
and construct the action of torsion coupled to spinor fields which
satisfies all the conditions required for the effective quantum theory.
Contrary to the other authors \cite{carroll} we take antisymmetric
torsion field to be parametrized by the massive vector field rather
than the scalar field because the theory of scalar torsion fails
to pass the test of invariant renormalizability. Then we incorporate
the Higgs scalar and discuss the violation of consistency which 
occurs due to the higher loop contributions. After discussion of
this problem and its possible solutions we present a phenomenological
consideration of the possible effects of the dynamical torsion coupled
to the fermions of the SM and
derive upper bounds for the torsion parameters from:\\
1) precisely measured forward-backward  Z pole asymmetries at LEP;\\
2) modern limits on the contact interactions coming from various low
   and high energy experiments;\\
3) TEVATRON data on search of the new vector bosons in di-jet channel.

The paper is organized in the  following way. In section 2
a brief review of the background notions for the gravity with torsion
is given and the form of interaction between torsion and other
fields is found. 
In section 3  the new specific gauge symmetry of
the massless part of the spinor action coupled to torsion is outlined.
Using this symmetry as a guide
we consider the renormalization of the spinor-torsion
system  and discuss the effective field theory approach to the
torsion phenomenology. It will be shown
that the new gauge symmetry requires the torsion
field to be (equivalent to) massive pseudovector. The explicit
calculation of the one-loop divergences confirms this result.
Section 4 is devoted to introducing the scalar field
and to related difficulties. In section 5 the renormalization group
equations for the torsion parameters are explored. We consider the
running of the torsion-spinor couplings and the torsion mass
and present some arguments that if the values of various 
spinor-torsion couplings are equal at 
the Planck scale, then those values are not very different at
the lower energies as well. Bearing this in mind we simply put these
parameter to be equal for all quarks and leptons and pass to the
phenomenological part of the paper.
In section 6 we discuss the possible physical observables
from which  an upper bounds
on  the torsion mass and fermion-torsion coupling can be derived.
In section 7  we analyze the limits
on torsion parameters coming from the LEP Z-pole forward-backward
lepton and quark asymmetries.
 Section 8 contains the 
discussion of a ``heavy'' torsion for which
the torsion phenomenology
can be successfully described by the axial-axial contact interactions.
 Section 9 contains the limits on the torsion parameters from
TEVATRON using  bounds on the cross section for di-jet events with
high invariant mass.
In the final section 10 we present our conclusions.

\vskip 6mm
\noindent
{\large\bf 2. Background notions of the gravity with torsion}
\vskip 2mm

Let us review the background notions for the gravity with torsion
and quantum theory of matter fields in the external torsion field.
All the notations correspond to  \cite{book} where one can also
find a more pedagogical introduction into the subject.

Consider the space - time with independent metric and torsion.
The affine connection
$\tilde{\Gamma}^\alpha_{\;\beta\gamma}$ is nonsymmetric, and
the torsion tensor $T^\alpha_{\;\beta\gamma}$  is defined as
\beq
{\tilde {\Gamma}}^\alpha_{\;\beta\gamma} -
{\tilde {\Gamma}}^\alpha_{\;\gamma\beta} =
T^\alpha_{\;\beta\gamma}
\label{tor}
\eeq
The covariant derivative $\tilde{\nabla}_\mu$ is based on the
nonsymmetric connection $\tilde{\Gamma}^\alpha_{\;\beta\gamma}$.
$\,$From the metricity condition
$\,\tilde{\nabla}_\mu g_{\alpha\beta} = 0\,$
the solution for the connection can be easily found in the form
\beq
\tilde{\Gamma}^\alpha_{\;\beta\gamma} =
{\Gamma}^\alpha_{\;\beta\gamma} + K^\alpha_{\;\beta\gamma}
\label{gam}
\eeq
where ${\Gamma}^\alpha_{\;\beta\gamma}$ is usual symmetric
Christoffel symbol and
$\,\,
K^\alpha_{\;\beta\gamma} = 
\frac{1}{2} \,(\, T^\alpha_{\;\;\beta\gamma} -
T^{\;\alpha}_{\beta\;\gamma} - T^{\;\alpha}_{\gamma\;\beta} \,)
\,\,$
is the contorsion tensor.
It proves useful to divide the torsion field into three irreducible
components:

i) $\,\,\,$ The vector (trace) 
$\,\,T_{\beta} = T^\alpha_{\;\beta\alpha}$;

ii) $\,\,$ The axial vector (pseudotrace)
$\;\,S^{\nu} = \varepsilon^{\alpha\beta\mu\nu}\,T_{\alpha\beta\mu}\;$

iii) $\,$ The tensor
$\;\,\,q^\alpha_{\;\beta\gamma}\;,\;\;\;\;
{\rm where}\;\;\;\;\;
q^\alpha_{\;\beta\alpha} = 0\;\;\;\;\;{\rm and}\;\;\;\;
\varepsilon^{\alpha\beta\mu\nu}q_{\alpha\beta\mu} =0$.

\noindent
In general case the torsion field can be presented in the form
\beq
T_{\alpha\beta\mu} = \frac{1}{3}\,\left(\, T_{\beta}\,g_{\alpha\mu} -
T_{\mu}\,g_{\alpha\beta}\, \right) - \frac{1}{6}\,
\vp_{\alpha\beta\mu\nu}\,S^{\nu} + q_{\alpha\beta\mu}
\label{gen}
\eeq

One can consider the Dirac
spinor $\psi$ in an external gravitational field with
torsion using both minimal and nonminimal schemes. The conventional
minimal way of introducing the interaction
with external fields is to substitute the partial
derivatives $\partial_\mu$ of spinors by the covariant 
${\tilde {\nabla}}_\mu$ ones. The covariant derivatives of the 
spinor field $\psi$ are defined as
\beq
\tilde{\nabla}_\mu \psi = \partial_{\mu}\psi +
\frac{i}{2}\,{\tilde{w}}_\mu^{\; a b}\,
\sigma_{a b}\,\psi
\,,\,\,\,\,\,\,\,\,\,\,\,\,\,\,\,\,\,\,\,\,\,\,\,
\tilde{\nabla}_\mu \bar{\psi} = \partial_{\mu}\bar{\psi} -
\frac{i}{2}\,{\tilde{w}}_\mu^{\;a b}\,\bar{\psi}\,\sigma_{a b}
\label{spinor}
\eeq
where $\tilde{w}_\mu^{\;a b}$ are the components of the spinor
connection, $\,\sigma_{a b} = \frac{i}{2}(\ga_a\ga_b - \ga_b\ga_a)\,,\,$
and we are using the standard representation for the Dirac matrices.
The $\,\gamma$-matrices in curved space-time are defined as $\gamma^\mu 
=
e_a^\mu \gamma^a$ where $e_a^\mu$ are the components of the verbein.
Indeed $\,\tilde{\nabla}_\mu \gamma^{\beta} = 0\,$. 
It is easy to find the explicit expression
for spinor connection which agrees with (\ref{gam}).
\beq
\tilde{w}_\mu^{\;a b} = 
\frac{1}{4} \,(\,e_\nu^b\, \partial_\mu e^{\nu a}
- e_\nu^a \,\partial_\mu e^{\nu b}\,) + 
{\tilde {\Gamma}}^\alpha_{\;\nu\mu}\,
(e^{\nu a}e_\alpha^b - e^{\nu b}e_\alpha^a)
\label{spicon}
\eeq
Substituting the
covariant derivatives (\ref{spinor}),
(\ref{spicon}) into the Hermitian action of the Dirac field
\beq
S = \frac{i}{2}\,
\int d^4 x \sqrt{-g}
\,\left\{ \,\bar{\psi}\gamma^\mu {\tilde {\nabla}}_\mu \psi
- {\tilde {\nabla}}_\mu\bar{\psi}\,\gamma^\mu\psi +
2im\,\bar{\psi}\psi\, \right\}
\label{herm}
\eeq
after some algebra we arrive at the following ``minimal'' action
\beq
S = \int d^4 x\sqrt{-g}\,\left\{
\,i\bar{\psi}\,\gamma^\mu\,(\partial_\mu+\frac{i}{8}\,\gamma_5\,
S_\mu\,)\psi\,+\,m\bar{\psi}\psi\,\right\}
\label{min}
\eeq
Below we will be interested only in the torsion effects and
therefore it is reasonable to restrict ourselves by the special case
of the flat metric. So we put $g_{\mu\nu} = \eta_{\mu\nu}$ but keep
$T^\alpha_{\;\beta\gamma}$ arbitrary. One can see that minimally the
spinor field interacts only with the pseudovector $S_\mu$ part of the
torsion tensor. The nonminimal interaction may be a bit more
complicated. Using the considerations based on dimensional reasons
one can introduce generic nonminimal coupling of the form
\beq
S = \int d^4 x\,\left\{
\,i\bar{\psi}\gamma^\mu\left(\partial_\mu + i \eta\gamma_5 S_\mu
- i {\hat {\eta}} T_\mu \right) \psi\,+\,m\bar{\psi}\psi\,\right\}
\label{nonmin}
\eeq

Despite the general nonminimal action (\ref{nonmin}) contains
two dimensionless nonminimal parameters
$\,\eta,{\hat {\eta}}\,$ we shall use only one of them
and put $\,{\hat {\eta}}=0\,$. There are several reasons to do so.
As we have already seen the minimal interaction includes only $\eta$
term. In the quantum theory of matter fields on an external
torsion background one meets, therefore, $\,\eta\,$ as an essential
parameter of the interaction while $\,{\hat {\eta}}=0\,$ is not
essential. In other words, if the $\,{\hat {\eta}}=0\,$-term is not
included into the classical action, the theory doesn't lose
renormalizability while fixing $\,\eta=\frac18\,$ such that interaction
is minimal leads to some difficulties 
\cite{bush1,book}. On the other hand, the
$\eta$-term looks very similar to the electromagnetic interactions.
If we introduce the interaction with the electromagnetic field then
the $\eta$-term can be revoked by simple redefinition of the 
variables and constants.
And, as a last reason we can remind that
in the string-induced action, which depends on the completely
antisymmetric torsion, only the pseudovector part
$\,S_\mu\,$ is present, and thus one can always set
$\,T_{\al\be\mu} = - \frac{1}{6}\,\ep_{\al\be\mu\nu}\,S^{\nu}\,.\,$
Below we always use the pseudovector $S_\mu$ to parametrize 
the completely antisymmetric torsion tensor.

With the scalar field $\,\ph\,$ torsion may interact only nonminimally,
because $\,{\tilde  {\na}}\ph = \pa\ph\,$. The action of free scalar
field including the nonminimal interaction with antisymmetric
torsion has the form
\beq
S_{sc} = \int d^4x\,\left\{\,\frac12\,g^{\mu\nu}\,\pa_\mu\ph\,\pa_\nu\ph
+\frac12\,m^2\,\ph^2 + \frac12\,\xi\,S_\mu S^\mu\,\ph^2\,\right\}
\label{scal}
\eeq
where $\,\xi\,$ is a new nonminimal parameter. If the quantum theory
contains scalar and spinor fields linked by the Yukawa interaction
$\,h\ph{\bar \psi}\psi\,$, then the nonminimal parameter $\xi$ is
necessary for the renormalizability. For the quantum field theory
in the external torsion field the renormalization of the
parameters $\,\eta,\xi\,$ possesses some universality. In particular,
the $\be$-function for the nonminimal parameter $\eta$ always has
the form
\beq
\be_\eta = \frac{C}{(4\pi)^2}\,h^2\,\eta\,,
\label{old}
\eeq
 where the value of the parameter
$C$ depends on the model but it is always positive \cite{bush2}.
In section 5 we shall see how the renormalization group equation for
$\eta$ modifies in case of the propagating torsion field.

We accept that the gauge vector field do not interact with torsion at
all, because such an interaction, generally, contradicts to the gauge
invariance. This can be easily seen from the relation
\beq
{\tilde \na}_\mu A_\nu - {\tilde \na}_\nu A_\mu =
{\pa}_\mu A_\nu - {\pa}_\nu A_\mu + K^\la_{\;\;\mu\nu}\,A_\la.
\label{vec}
\eeq
The nonminimal interaction with abelian vector field may be indeed
implemented in the form of the surface term
\beq
S_{n-m,vec}\,=
\,i\,\al\,\int d^4x\,\ep^{\al\be\ga\si}\,F_{\al\be}\,S_{\ga\si}.
\label{surf}
\eeq
Other nonminimal terms \cite{book} are also possible for the general
torsion but they are relevant only for the nonzero
$T_\mu$ and $q_{\al\be\ga}$ of the torsion tensor \cite{bush2} and 
thus we are not interested in them here.

\vskip 6mm
\noindent
{\large\bf 3. Effective approach in the spinor-torsion systems}
\vskip 2mm

Our aim it to formulate the effective quantum theory of torsion 
such that it would be
compatible with the requirements of renormalizability and unitarity.
Let us first consider this problem for the system without scalar fields.
As it was already noticed the renormalizability of the gauge model
in an external torsion field requires the nonminimal interaction of the
spinor fields with torsion \cite{bush1}.
It will prove reasonable to introduce such an interaction
in the theory with the propagating torsion too. Thus we start from the
action of the Dirac spinor nonminimally coupled to the electromagnetic
and torsion fields
\beq
S_{1/2}= i\,\int d^4x\,{\bar \psi}\, \left[
\,\ga^\al \,\left( {\pa}_\al - ieA_\al + i\,\eta\,\ga_5\,S_\al\,\right)
- im \,\right]\,\psi
\label{diraconly}
\eeq
and first establish its symmetries.
The new interaction with torsion doesn't spoil
the invariance of the above action under usual gauge transformation:
\beq
\psi' = \psi\,e^{\al(x)}
,\,\,\,\,\,\,\,\,\,\,\,\,\,\,
{\bar {\psi}}' = {\bar {\psi}}\,e^{- \al(x)}
,\,\,\,\,\,\,\,\,\,\,\,\,\,\,
A_\mu ' = A_\mu - {e}^{-1}\, \pa_\mu\al(x)
\label{trans1}
\eeq
It turns out, however, that there is one more symmetry.
The massless  part of  the action (\ref{diraconly})
is invariant under the transformation in which the pseudotrace of
torsion plays the role of the gauge field
\beq
 \psi' = \psi\,e^{\ga_5\be(x)}
,\,\,\,\,\,\,\,\,\,\,\,\,\,\,
{\bar {\psi}}' = {\bar {\psi}}\,e^{\ga_5\be(x)}
,\,\,\,\,\,\,\,\,\,\,\,\,\,\,
S_\mu ' = S_\mu - {\eta}^{-1}\, \pa_\mu\be(x)
\label{trans}
\eeq
Thus in the massless sector of the theory one faces generalized
gauge symmetry depending on scalar $\al(x)$ and pseudoscalar
$\be(x)$ parameters of transformation. It is very important that the
massive term is not invariant under the transformation (\ref{trans}),
and hence this symmetry is softly broken at classical level.
The new symmetry (\ref{trans}) requires $S_\mu$ to be massive 
field and fixes the action of torsion with
accuracy to the values of the nonminimal parameter $\eta$, mass of the
torsion $M_{ts}$ and possible higher derivative terms.
The mass of the torsion is necessary because the softly broken
symmetry (\ref{trans}) doesn't forbid the appearance of the
massive counterterms (contrary to the situation for the abelian gauge
field). We shall give some more arguments in favor
of massive torsion in section 5 after discussion of the
renormalization group equations. One more inconsistency related with
the massless torsion comes from the calculation of the anomalous
magnetic moment of the electron, which has been performed recently
\cite{kalm}. Contrary to the abelian vector field, massless axial
field leads to the IR divergency.

In the framework of effective field theory the contributions from the
loops of a very
massive fields are suppressed by the factors of $\,\mu^2 / M^2$ where
$M$ is the mass of the field and $\mu$ the typical energy of the
process \cite{apecor}. If we take the torsion mass
$M_{ts}$ to be of the Planck order then both classical and quantum
effects of torsion will be negligible at the energies available at the
modern experimental facilities. The hypothesis of torsion propagating
at energies lower than the Planck one
supposes that $M_{ts}$ is essentially smaller than the Planck mass. Then
we have two options: take torsion to be massless or consider the mass
of torsion as a free parameter which should be defined on an
experimental basis. As far as torsion is taken as a dynamical field,
one has to incorporate it into the SM along with other vector fields.
Let us discuss the form of the torsion action, which leads to the
consistent quantum theory. The higher derivative terms are supposed
to be included into the action, but they are not seen at low energies 
and thus have no importance.
Thus we restrict the torsion action by the second
derivative and zero-derivative terms.
The general action including these terms has the following form:
\beq
S_{tor} = \int d^4\,\left\{\, -a
\,S_{\mu\nu}S^{\mu\nu} + b\,(\pa_\mu S^\mu)^2
+ M_{ts}^2\,S_\mu S^\mu\,\right\}
\label{geral}
\eeq
where $\,S_{\mu\nu} = \pa_\mu S_\nu - \pa_\mu S_\nu\,$ and $a,b$
are some positive parameters. The action (\ref{geral}) contains both
transversal vector mode and the longitudinal mode which is in fact
equivalent to the scalar\footnote{This kind of torsion equivalent to the
pseudoscalar field was introduced in \cite{novello} in order to
maintain the gauge invariance of abelian vector
field in the Riemann-Cartan spacetime.}
(see, for example, discussion in \cite{carroll})
In particular, in the $a=0$ case only the scalar mode,
and for $b=0$ only the vector mode propagate.
It is well known \cite{vector} (see also \cite{carroll} for the
discussion of the theory (\ref{geral})) that in the unitary
theory of the vector field both longitudinal and transversal
modes can not propagate, and therefore, in order to have consistent
theory of torsion one has to choose one of the parameters $a,b\,$
to be zero.

In fact the only correct choice is $b=0$. To see this one has to reveal
that the symmetry (\ref{trans}), which is spoiled by the massive terms
only, is
always preserved in the renormalization of the dimensionless couplings
constants of the theory. In other words, the divergences and
corresponding local counterterms, which produce the dimensionless
renormalization constants, do not depend on the dimensional parameters
such as the masses of the fields. This structure of renormalization is
essentially the same as for the Yang-Mills theories with spontaneous
symmetry breaking \cite{wein,votyu}. The symmetry (\ref{trans}) holds
for the massless part of the action (\ref{diraconly}) and therefore
on the dimensional grounds one has to expect that the gauge invariant
counterterm $\,\,\int S_{\mu\nu}^2\,\,$ appears if we take the loop
corrections into account.

We want emphasize that in the framework of effective field
theory the level of approximation for taking into
account the massive fields is qualitatively the same
for the tree level and
for the lower loop effects. Since the propagating
torsion is considered and the kinetic term in (\ref{geral})
is taken into account, one has to formulate the theory
as renormalizable. Neglecting the high energy effects while the
low energy amplitudes are considered may mean that we disregard
some higher derivative terms. However the violation of the
renormalizability
in that sectors of the theory which are taken into account is 
impossible.
For instance, if we start from the purely scalar longitudinal torsion
(as the authors of \cite{carroll} did) then the transversal term
$\,\int S_{\mu\nu}^2\,$ will arise with the divergent coefficient
and this will indeed violate both the finiteness of the effective
action and the unitarity of the $S$-matrix. All this is true even
in the case that only the tree-level effects are evaluated, if only such
consideration is regarded as an approximation to any reasonable
quantum theory.

Thus the kinetic term of the torsion action is given by the
Eq. (\ref{geral}) with $b=0$. As concerns the massive term it is not
forbidden by the symmetry (\ref{trans}), because the last is softly
broken. 
Therefore apriory there are no reasons to suppose that $M_{ts}=0$.
Indeed it is interesting to see whether the kinetic counterterm
with $b=0$ and the massive counterterm really
appear if we take into account the fermion loops. To investigate this
let us calculate the one-loop divergences in the
theory (\ref{diraconly})
\footnote{Similar calculation for the massless theory in curved
space-time has been performed in \cite{buodsh}.}.
The divergent part of the one-loop effective action is given by the
expression
\beq
\Ga_{div} [A, S] = - \Tr\ln {\hat H}\mid_{\,div}\,;
\,\,\,\,\,\,\,\,\,\,\,\,\,\,\,\,\,
{\hat H} =
i\ga^\al \,\left( {\pa}_\al - ieA_\al + i\eta\ga_5\,S_\al\,\right)-im
\label{efac}
\eeq
In order to calculate this functional determinant one has to perform the
transformation
\beq
 \Tr\ln {\hat H}
=  \Tr\ln {\hat H}\cdot \left(i\ga^\nu\pa_\nu\right)
-  \Tr\ln \left(i\ga^\nu\pa_\nu\right) =
 \Tr\ln \left(-{\hat 1}{\Box} - 2{\hat h}^\nu\pa_\nu\right)
-  \Tr\ln \left(i\ga^\nu\pa_\nu\right)
\label{kvadrat}
\eeq
with
$$
{\hat h}^\nu = \frac{i}{2}\,\left(\, -\,e\,A_\mu\,\ga^\mu\ga^\nu
\,-\,\eta\,S_\mu\,\ga^\mu\ga_5\ga^\nu + m\ga^\nu\,\right)
$$
The second term in the right-hand side of Eq. (\ref{kvadrat}) is
a constant which doesn't depend on $S_\mu$ or $A_\mu$, while the
first term can be easily evaluated using the standard Schwinger-deWitt
technique (one can see \cite{book} for the introduction and references).
After some algebra we arrive at the following counterterm:
\beq
\De S [A_\mu, S_\al] = \frac{\mu^{n-4}}{\varepsilon} \int d^\mu x
\,\left\{ \frac{2e^2}{3}F_{\mu\nu}F^{\mu\nu} +
\frac{2\eta^2}{3}S_{\mu\nu}S^{\mu\nu}
- \frac{ie\eta}{3}\ep^{\al\be\mu\nu}S_{\mu\nu}F_{\al\be}
+ 8m^2\eta^2S^\mu S_\mu  \right\}
\label{contra}
\eeq
where $\,\varepsilon = (4\pi)^2\,(n-4)\,$ is the parameter of
the dimensional regularization.
Here we disregarded all surface terms except the 
$\ep^{\al\be\mu\nu}\,S_{\mu\nu}F_{\al\be}$ one, because it can, in
principle, lead to quantum anomaly. It would be interesting to 
explore this possibility but since the phenomenological consideration
below is mainly restricted by the tree-level effects this
term is beyond the scope of our present interest. The form of the
counterterms (\ref{contra}) is in perfect agreement with the
above analysis based on the symmetry transformation (\ref{trans}).
Namely, the one-loop divergences contain $S_{\mu\nu}^2$ and 
the massive
term while the $\,\left(\pa_\nu S^\nu\right)^2$ term is absent.

Thus the correct form of the torsion action which can be coupled
to the spinor field (\ref{diraconly}) and lead to the unitary and
renormalizable theory is
\beq
S_{tor} = \int d^4\,\left\{\, -\frac14\,S_{\mu\nu}S^{\mu\nu}
+ M_{ts}^2\, S_\mu S^\mu\,\right\}\,.
\label{action}
\eeq
In the last expression we put the conventional coefficient $\,-1/4$
in front of the kinetic term. With respect to the renormalization
this means that we (in a direct analogy
with QED) can remove the kinetic counterterm by the renormalization
of the field $S_\mu$ and then renormalize the parameter $\eta$ in
the action (\ref{diraconly}) such that the combination
$\eta S_\mu$ is the same for the bare and renormalized quantities.
Instead one can include $1/\eta^2$ into the kinetic term of
(\ref{action}), that should lead to the direct renormalization of
this parameter while the interaction of torsion with spinor has
minimal form (\ref{min}) and $S_\mu$ is not renormalized.
Therefore in the case of propagating torsion the difference
between minimal and nonminimal types of interactions is only the
question of notations on both classical and quantum levels.

\vskip 6mm
\noindent
{\large\bf 4. Introducing scalar field}
\vskip 2mm

Despite the Higgs scalar is not detected experimentally, it is
considered as an important constituent of the SM. The introduction
of the scalar field is necessary for the spontaneous symmetry breaking
and for the Higgs mechanism which makes the gauge bosons massive
and enables one to avoid the infrared divergences in gauge theories.
As far as we are going to incorporate torsion into the SM it is
important to extend our consideration introducing scalar field and
Yukawa interactions. Doing this we shall follow the
same line as in the previous section and try first to construct the
renormalizable theory. Hence the first thing to do is to analyze
the structure of the possible divergences. The divergent diagrams
in the theory with a dynamical torsion include, in particular, all
such diagrams with external lines of torsion and internal lines
of other fields. Those grafs are indeed the same one meets in the
quantum field theory on an external torsion background. Therefore
one has to include into the action all the terms which were necessary
for the renormalizability when torsion was purely external field.
All such terms are well-known from \cite{bush1,bush2}. Besides the
nonminimal interaction with spinors
one has to introduce the nonminimal interaction between scalar field
and torsion as in (\ref{scal}) and also the terms which played the
role of the action of vacuum (see chapter 4 of \cite{book} where the
one-loop counterterms for scalar field are also presented) in the form
\beq
S_{tor} = \int d^4\,\left\{\, - \frac14\,S_{\mu\nu}S^{\mu\nu}
+ M_{ts}^2\, S_\mu S^\mu - \frac{1}{24}\,\ze \left(S_\mu S^\mu\right)^2
\,\right\} + {\rm surface\, terms}\,.
\label{act}
\eeq
Here $\,\ze\,$ is new arbitrary parameter, and coefficient
$\frac{1}{24}$ stands for the sake of convenience only. And so, if one
introduces torsion into the whole SM including the scalar field, the
total action includes the following new terms: action of torsion
(\ref{act}) with the self-interacting term, and a nonminimal
interactions between torsion and spinors (\ref{diraconly}) and scalars
(\ref{scal}). It is easy to see that such a theory suffers from a 
serious difficulty.

\begin{figure}[htb]
   \begin{center}
    \vskip -3cm\hspace*{-3cm}
    \epsfxsize=12cm\epsffile{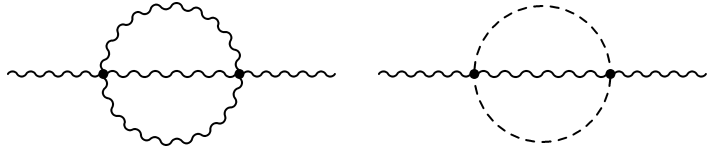}
    \vskip -12cm
\caption{\it The
wavy line is torsion propagator and dashed line -- scalar propagator}
\label{fig:dang}
\end{center}
\end{figure}

The root of the problem is that the Yukawa interaction term
$\,h\ph{\bar {\psi}}\psi$ is not invariant under the transformation
(\ref{trans}). Unlike the spinor mass the Yukawa constant $h$ is
massless, and therefore this noninvariance may affect the
renormalization in the massless sector of the theory. In particular,
the noninvariance of the Yukawa interaction causes the necessity
of the nonminimal scalar-torsion interaction in (\ref{scal}) which,
in turn, requires an introduction of the self-interaction term in
(\ref{act}). Those terms do not pose any problem at the one-loop
order but already in the second loop one meets two dangerous diagrams
presented in Fig.~\ref{fig:dang}

Those diagrams are divergent and they can lead
to the appearance of the $\,\left(\pa_\mu S^\mu\right)^2\,$-type
counterterm. No any symmetry is seen which forbids these divergences.
Let us consider two diagrams presented on Fig.1 in more details.
Using the
actions of the scalar field coupled to torsion (\ref{scal}) and
the torsion self-interaction (\ref{act}), we arrive at the following
Feynman rules:
\vskip 2mm

\noindent
i) Scalar propagator: $\,\,\,\,\,\,\,\,\,
\,\,\,\,\,\,G(k) = \frac{i}{k^2+M^2}\,\,{\rm where}\,\,\,M^2=2M_{ts}^2$,
\vskip 1mm

\noindent
ii) Torsion propagator:  $\,\,\,\,\,\,
\,\,\,\,\,\,D_\mu^{\,\nu}(k) = \frac{i}{k^2+M^2}
\,\left(\de_\mu^{\,\nu} + \frac{k_\mu k^{\nu}}{M^2} \,\right)\,\,,$
\vskip 1mm

\noindent
iii) Torsion$^2$--scalar$^2$ vertex: $\,\,\,\,\,\,\,\,\,\,\,\,
V^{\mu\nu}(k,p,q,r)\, = \,- \,2 i\xi\,\eta^{\mu\nu}\,\,,$
\vskip 1mm

\noindent
iv) Vertex of torsion self-interaction: 
$\,\,\,\,\,\,V^{\mu\nu\al\be}(k,p,q,r)\, =
 \,\frac{i\ze}{3}\,g_{(2)}^{\mu\nu\al\be}$
\vskip 2mm

\noindent
where $g_{(2)}^{\mu\nu\al\be} = g^{\mu\nu}g^{\al\be} +
g^{\mu\be}g^{\al\nu} + g^{\mu\al}g^{\nu\be}$ and $k,p,q,r$ denote the
outgoing momenta.

The only one thing that we would like to check is the violation of 
the transversality in the kinetic 2-loop counterterms.
We shall present the calculation in some details because it is
quite instructive.
To analize the loop integrals we have used dimensional
regularization and in particular the formulas from \cite{hela}.
It turns out that it is sufficient to trace the
$\frac{1}{\vp^2}$-pole, because even this leading divergency
requires the longitudinal counterterm.
The contribution to the mass-operator of torsion from the second
diagram from Fig.1 is given by the following integral
\beq
\Pi^{(2)}_{\al\be}(q)\,=\, -\,2\, \xi^2\,
\int \frac{d^{D} k}{(2\pi)^4} \frac{d^{D} p}{(2\pi)^4}
\frac{  \,\eta_{\al\be}\, + \,M^{-2}\, (k-q)_\al\,(k-q)_\be \,}{(p^2+
M^2)[(k-q)^2+M^2][(p+k)^2 + M^2]}
\label{int1}
\eeq
To perform the integration one can expand one of the factors in the
above integral in the following way:
\beq
\frac{1}{(k-q)^2+M^2} =
\frac{1}{k^2+M^2}\,\left[1 + \frac{-2k\cdot q+q^2}{k^2+M^2}\right]^{-1}=
\frac{1}{k^2+M^2} \,
\sum_{n=1}^{\infty}(-1)^n\,
\left(\frac{-2k\cdot q+q^2}{k^2+M^2}\right)^{n}
\label{expand}
\eeq
and substitute this expansion into (\ref{int1}). It is easy to see that
the divergences hold in this expansion till the order $n=8$. On the
other hand, each order brings some powers of the external momenta
$q^\mu$. Therefore the divergences of the above integral may be
cancelled by adding the counterterms which include high derivatives.
To achieve the renormalizability one has to include these high 
derivative
terms into the action (\ref{act}). 
However, since we are aiming to construct the effective (low-energy)
field theory of torsion, the effects of the higher derivative terms are
not seen and their renormalization is not interesting for us. All we 
need are the second derivative counterterms. Hence, for our purposes
the expansion (\ref{expand}) can be cut at $n=2$ rather that at $n=8$
and moreover only $O(q^2)$ terms should be kept. Then,
using also usual symmetry considerations, one arrives at the
known (see, for example, \cite{hela}) intergal
$$
\Pi^{(2)}_{\al\be}(q) = -6\xi^2\,q^2\,\eta_{\al\be}\,
\int \frac{d^{D} k}{(2\pi)^4}
\frac{d^{D} p}{(2\pi)^4}\,\frac{1}{p^2+M^2}
\,\frac{1}{(k^2+M^2)^2}\,\frac{1}{(p+k)^2+M^2} + \,...\,=
$$
\beq
= -\frac{12\,\xi^2}{(4\pi)^4\,(D-4)^2}\,q^2\,\eta_{\al\be} 
\,+\,({\rm lower}\,\,{\rm poles})\, + \, ({\rm higher} \,\, {\rm
derivative} \,\, {\rm terms}).
\eeq
Another integral looks a bit more complicated, but its derivation
performs in a similar way.
The contribution to the mass-operator of torsion from the first
diagram from Fig.1 is given by the integral
$$
\Pi^{(1)}_{\al\la}(q)=- \frac{\ze^2}{108}\,
g^{(2)\al\rho\si\be}\,g^{(2)\la\tau\mu\nu}\,
\int \frac{d^4 k}{(2\pi)^4}\,\int \frac{d^4 p}{(2\pi)^4}\,
\frac{1}{k^2+M^2}
\left(\eta_{\tau\be} + \frac{k_\tau k_\be}{M^2}\right)\times
$$
\beq
\times
\left(\eta_{\rho\mu} +
\frac{(p-q)_\rho (p-q)_\mu}{M^2}\right)\,\frac{1}{p^2+M^2}
\,\left( \eta_{\si\nu} +
\frac{(p+k)_\si (p+k)_\nu }{M^2}\right)\,\frac{1}{(k+q)^2+M^2 }
\label{int3}
\eeq
Now, we perform the same expansion (\ref{expand}) and, disregarding
lower poles, finite contributions and higher derivative leading
divergences arrive at
\begin{eqnarray}
\Pi^{(1)}_{\al\la}(q)&=&- \frac{\ze^2}{108}\,
g^{(2)\al\rho\si\be}\,g^{(2)\la\tau\mu\nu}\,\times \nonumber\\
&& \times 3\, q^2\,\eta_{\tau\be}\,\eta_{\rho\mu}\,\eta_{\si\nu}\, 
\int \frac{d^4 k}{(2\pi)^4}\,\int \frac{d^4 p}{(2\pi)^4}\,
\frac{1}{k^2+M^2}\frac{1}{(p^2+M^2)^2}\frac{1}{(k+p)^2+M^2}
\,+\, ...\,.
\end{eqnarray}
After a simple algebra this leads to the following leading divergency
\beq
\Pi^{(1)}_{\al\la}(q)=
- \frac{\ze^2}{(4\pi)^4\,(D-4)^2}\,q^2\,\eta_{\al\la} +\, ...\,.
\label{int4}
\eeq

Thus we see that both diagrams from Fig.1 really give rise to the
longitudinal kinetic counterterm and no any simple
cancellation of these divergences is seen. On the other hand
one can hope to achieve
such a cancellation on the basis of some sofisticated symmetry. 

To understand the situation better let us compare it with the one that
takes place for the usual abelian gauge transformation
(\ref{trans1}). In this case the symmetry is not violated by the
Yukawa coupling, and (in the abelian case) the $\,A^2\ph^2\,$
counterterm is impossible because it violates gauge invariance.
The same concerns also the self-interacting $A^4$ counterterm.
The gauge invariance of the theory on quantum level is controlled
by the Ward identities.
In principle, the noncovariant counterterms can show up, but they
can be always removed, even in the non-abelian case,
by the renormalization of the gauge parameter
and in some special class of (background) gauges they are impossible
at all. Generally, the renormalization can be always performed in a
covariant way \cite{volatyu}.

In case of the transformation (\ref{trans})
if the Yukawa coupling is inserted there are no reasonable gauge
identities at all. Therefore in the theory of torsion field coupled
to the SM with scalar field there is a conflict between
renormalizability and unitarity. The action of the renormalizable
theory has to include the $\,\left(\pa_\mu S^\mu\right)^2\,$ term,
but this term leads to the problems with the positivity of the energy
and, in terms of particles, to the appearance of the massive ghost.
This conflict between unitarity and renormalizability reminds the
one which is well known -- the problem of massive unphysical ghosts
in the high derivative gravity \cite{stelle}, where the
 contributions of the massive ghosts provide renormalizability
but breaks the unitarity of the
$S$-matrix. The difference is that in our case, unlike higher
derivative gravity,
the problem appears due to the noninvariance with respect
to the transformation (\ref{trans}).

Let us now discuss how this problem may be, in principle, solved.
First thought is that if the torsion mass is of the Planck order then
the quantum effects of torsion should be described directly in the
framework of string theory. No any effective field theory for torsion
is possible. In this case the only visible term in the torsion action
is the massive one in (\ref{act}), torsion does not propagate at
smaller energies and manifests itself only as a very weak contact
interaction. We shall give the discussion of these contact interactions
below in section 7, however the phenomenological
analysis fails for $M_{ts}$ well below the Planck order.

There may be a hope to impose one more symmetry which is
not violated by the Yukawa coupling. It can be, for example,
supersymmetry which mixes torsion with some vector fields of the
SM and with all massive spinor fields. In this case the
$\,\left(\pa_\mu S^\mu\right)^2\,$-type counterterm may be
forbidden by this symmetry and the conflict between renormalizability
and unitarity would be resolved.

Another option is to consider the modification of SM which is free
from the fundamental scalar fields at all. 
This possible scenario is related with the fact that
there is one crucial point of SM which still remains unclear:
the pattern of electroweak symmetry breaking which generates masses
for the $W$ and $Z$ bosons. Moreover, we still lack the same accuracy
tests for the triple and quartic bosonic interactions to further
confirm the local gauge invariance of the theory or to indicate the
existence of new physics beyond the SM.

The interactions responsible for the electroweak symmetry breaking
play an important role in the gauge--boson interactions at the TeV
scale since it is an essential ingredient to avoid unitarity violation
in the scattering amplitudes of massive vector bosons at energies of
the order of 1 TeV \cite{unit}.  There are two possible forms of
electroweak symmetry breaking sector which lead to different solutions
to the unitarity problem: there is, either, a scalar particle lighter
than 1 TeV, the Higgs boson of the Standard Model (SM), or such
particle is absent and the longitudinal components of the $W$ and $Z$
bosons become strongly interacting at the energy scale of 1 TeV.  In
the latter case, the symmetry breaking occurs due to the nonzero
vacuum expectation value of some composite operators, and it is
related with some new underlying physics.
Then the spontaneous
symmetry breaking can be realized through some other object like
composite scalar field or maybe through the torsion itself.

In the future sections of the paper we are going to discuss a
possible consequences
of the torsion action at low energies (as compared to the Planck one)
and find some numerical upper bounds for the parameters of this action.
Since the interaction with the scalar field leads to serious problems,
we shall consider only the interaction between torsion and spinor
field, and therefore use (\ref{action}) as the torsion action.
Thus we admit that one of the two last options (or some other)
for the resolution of the problem with the scalar field may be
successfully realized and that $M_{ts}$ is much less that the Planck
mass. So, instead of taking the string-inspired mass $M_{Pl}$
we take $M_{ts}$ to be some free parameter of the theory, as we do
with a couplings $\eta$. Indeed the last may be different for various
quarks and leptons.

\vskip 6mm
\noindent
{\large\bf 5. Renormalization group and the running of torsion mass and
parameters}
\vskip 2mm

In this section we shall discuss the renormalization group in the
theory with torsion. First we consider the spinor-torsion system
with an additional electromagnetic field, but without the
controversial scalar. Then the renormalization group equations for
the parameters $\,e,\eta,m,M_{ts},\al$ (here $\al$ is the parameter
of the nonminimal term (\ref{surf})) follow from Eq. (\ref{contra}).
\beq
(4\pi)^2\,\frac{de}{dt} = \frac23\,e^2
\,,\,\,\,\,\,\,\,\,\,\,\,\,\,\,\,\,\,\,\,\,   e(0) = e_0
\label{e}
\eeq
\beq
(4\pi)^2\,\frac{d\eta}{dt} = \frac23\,\eta^2
\,,\,\,\,\,\,\,\,\,\,\,\,\,\,\,\,\,\,\,\,\,   \eta(0) = \eta_0
\label{eta}
\eeq
\beq
(4\pi)^2\,\frac{d\al}{dt} = \frac13\,e\,\eta
\,,\,\,\,\,\,\,\,\,\,\,\,\,\,\,\,\,\,\,\,\,   \al(0) = \al_0
\label{regr}
\eeq
\beq
(4\pi)^2\,\frac{dM_{ts}^2}{dt} = 8\,m^2\,\eta^2 - 2\,M_{ts}^2
\,,\,\,\,\,\,\,\,\,\,\,\,\,\,\,\,\,\,\,\,\,   M_{ts}(0) = M_{ts,0}
\label{mass}
\eeq
We remark that the equation (\ref{mass}) demonstrates the inconsistency
of the massless or very light torsion. Even if one imposes the
normalization condition $M_{ts,0}\approx 0$ at some scale $\mu$, the
first term in this equation provides a rapid change of $M_{ts}$
such that it will be essentially nonzero at other scales. Due to the
universality of the interaction (\ref{diraconly})
all quarks and massive leptons should contribute to this
equation. Therefore the only way to avoid an unnaturally fast
running of $M_{ts}$ is to take its value at least of the order
of the heaviest spinor field that is $t$-quark. Hence we have some
grounds to take $M_{ts} \geq 100\,$GeV. Of course there can not be any
upper bounds for $M_{ts}$ from the equation (\ref{mass}).

The solution of the equations (\ref{e}) -- (\ref{regr}) for the
dimensionless parameters is simple. After the simple calculus we get
\beq
e(t) = e_0\,\left(1 - \frac{2\,e_0^2\,t}{3\,(4\pi)^2}\right)^{-1}
\,,\,\,\,\,\,\,\,\,\,\,\,\,\,\,\,\,
\eta(t) = \eta_0\,\left(1 - 
\frac{2\,\eta_0^2\,t}{3\,(4\pi)^2}\right)^{-1}\,,
\label{sole}
\eeq
\beq
\al(t) =  A \, + \,\frac{3\,e_0\,\eta_0}{2\,(e_0 - \eta_0)}
\ln \left( \frac{1 - \frac{2\,\eta_0^2}{3(4\pi)^2}\,t}{1 -
\frac{2\,e_0^2}{3(4\pi)^2}\,t}\right)
\,,\,\,\,\,\,\,\,\,\,\,\,\,\,\,\,\,
e_0 \neq \eta_0\,\,,\,\,\,\,\,\,\,\,\,\,\,\,\,\,A = const.
\label{soleta}
\eeq
The behavior of the effective couplings $\,e(t),\eta(t)\,$ is
usual for the nonabelian vector fields, and the running of $\eta$
is universal for all the spinors. The zero-charge problem doesn't
impose real restrictions on $\eta$ because for $\,\eta< 1 \, $
the singular point of the solution is many orders bigger than
$M_{Pl}$.

In order to see how the difference between various spinor fields
may arise let us consider the theory with the scalar fields and
Yukawa coupling but in the one-loop approximation where the formal
problems discussed in the previous section do not show up. At the
one-loop level the $\be$-functions for the different $\eta$'s come
from the renormalization of the kinetic term for torsion
of the $\,\,{\bar \psi}\ga_5\ga^\mu S_\mu\psi$-type. The last ones are
the same as for the quantum field theory in an external torsion
field. Taking into account
(\ref{old}) the full renormalization group equation for $\eta(t)$
is written as
\beq
(4\pi)^2\,\frac{d\eta}{dt} = \frac23\,\eta^2 + C\,h^2(t)\,\eta
\,,\,\,\,\,\,\,\,\,\,\,\,\,\,\,\,\,\,\,\,\,   \eta(0) = \eta_0
\label{eta1}
\eeq
One can use (\ref{eta1}) to evaluate the difference between the
running of $\,\eta\,$ for different spinor fields. Let us take,
for simplicity, constant $h$. Then the solution of (\ref{eta1})
has the form
\beq
\frac{\eta(t)}{\eta_0} = \frac{3Ch}{2}\,
\left[ \,( \eta_0 + \frac{3Ch}{2}\,) \cdot
exp\left\{ -\frac{Cht}{(4\pi)^2} \right\} - \eta_0 \right]^{-1}
\label{resh1}
\eeq
If we suppose that the value of $\eta_0$ is very small, then the
terms containing $\eta_0$ in the right-hand side can be abandoned
and we arrive at the following approximate solution
\beq
\frac{\eta(\mu)}{\eta_0} =
\left[\frac{\mu}{\mu_0}\right]^{\frac{Ch}{(4\pi)^2}}
\label{resh}
\eeq
Taking the normalization scale $\mu_0 = 100\,$GeV and the high-energy
scale $\mu=M_{Pl}=10^{19}$GeV, we find that for the $t$-quark
Yukawa constant 
$h_t\approx 0.98$ this ratio is about $10^{0.105\cdot C}$.
The value of $C$ depends on the gauge group and representation
which we do not know for all the energy ranges between $\mu_0$ and
$M_{Pl}$. Taking, for instance, the adjoint representation of the
$SU$(5) group we meet $C = 5$ \cite{book},
and therefore the ratio (\ref{resh})
is about $3.4$ for the $t$-quark while it is close to one for the
light spinors which have small Yukawa couplings.
We note that taking into account the running of $h(t)$ the resulting
ratio becomes a bit smaller.

Suppose the spinor fields are generated from some fundamental theory
at the Planck scale and originally have universal interaction with
torsion, with an equal parameters $\eta$. Then the difference in the
values of $\eta$ for various fields should be caused by the running.
As we have just seen, in the framework of this (essentially one-loop)
model these parameters still have the same order of magnitude at the
Fermi scale. Below we consider the processes which involve only one
type of the spinor fields, and hence for our purposes it is not
so important whether various $\eta$'s have equal value. However
this information will be implicitly used when we construct the general
limits for the torsion parameters from various known experiments.

\vskip 6mm
\noindent
{\large\bf 6. Phenomenology of torsion}
\vskip 2mm

As we have already seen, the spinor-torsion interactions
enter the Standard Model as interactions of fermions with 
new axial vector field $S_{\mu}$. 
Such an interaction is characterized by the
new dimensionless parameter  -- coupling constant $\eta$. Furthermore
the mass of the torsion field $M_{ts}$ is unknown, and its
value is of crucial importance for the possible experimental
manifestations of the propagating torsion and finally for the existence
of torsion at all. In this and consequent sections we consider
$\,\eta\,$ and $\,M_{ts}\,$ as an arbitrary parameters and try to limit 
their
values from the known experiments. 
Indeed we use the
renormalization group as an insight concerning the mass of torsion
but include the discussion of the "light" torsion with the mass of
the order of 1 GeV for the sake of generality. 


Our strategy will be to use known experiments directed to the
search of the new interactions. We regard torsion as one of those
interactions and obtain the limits for the torsion parameters from
the data which already fit with the phenomenological considerations.
Therefore in the course of our work we insert torsion
into the minimal SM and suppose that the other possible new physics is 
absent. It is common assumption when one wants to put limits
on some particular kind of a new physics.
In the following sections we put the limits on the parameters of the
torsion action using results of various experiments. 

Torsion, being a
pseudo-vector particle interacting with fermions might give therefore
different physical observables.
The main feature of torsion is related with its axial vector type 
interaction with fermions. This specific type of interaction
might lead to the forward-backward asymmetry. The last has been 
presizely measured at the LEP $e^+e^-$ collider, so the upper 
bounds for torsion
parameters may be set from  those measurements. 
We will  consider two different cases: i) torsion is much more
heavy than other particles of SM and $\,\,$
ii) torsion has a mass comparable to that
of other particles. In the last case one meets a propagating
particle which
must be treated on an equal footing with other constituents of the SM.
Contrary to that, the very heavy torsion leads to the effective contact
four-fermion interactions.

Consider the case of heavy torsion in some more details, starting
from the actions (\ref{diraconly}) and (\ref{action}). Since
the massive term dominates over the covariant kinetic
part of the action, the last can be disregarded. Then the total 
action leads to the algebraic equation of motion
for $\,S_\mu$. The solution of this equation can be substituted back to
$\,S_{1/2} + S_{tor}\,$ and thus produce the contact four-fermion
interaction term
\beq
{\cal L}_{int} = - \frac{\eta^2}{M_{ts}^2}\,
({\bar \psi}\ga_5\ga^\mu\psi)\,({\bar \psi}\ga_5\ga_\mu\psi)
\label{contact}
\eeq

As one can see the only one quantity which appears in this 
approach is the ratio ${M_{ts}}/{\eta}\,$ and therefore for the very 
heavy torsion field the phenomenological consequences depend only 
on single parameter.

Physical observables related with torsion depend on the two 
parameters $M_{ts}$ and $\eta$.
In the course of our study we choose, for the sake of simplicity,
all the torsion couplings with fermions to be the same $\eta$.
This enables one to put the limits in the two
dimensional ($M_{ts}$-$\eta$) parameter space using
the present experimental data.
We also assume that non-diagonal coupling of the torsion with the  
fermions 
of different families
is zero in order to avoid flavor changing neutral current problem.

It should be stressed that all numerical and symbolic calculations
for establishing limits on torsion parameters have been done using
CompHEP software package \cite{comp} to which the torsion
propagator and vertices were additionally introduced.

\vskip 10mm
\noindent
{\large\bf 7. Limits on the torsion parameters from presize
LEP electroweak data}
\vskip 2mm

As we mentioned above, the
axial-vector type interactions would give rise to the forward-backward
asymmetry which have been presizely measured in the
$e^+e^- \rightarrow l^+l^- (q\bar{q})$  scattering
(here $l$ stands for tau,muon or electron) at LEP collider with the
center-mass energy equals to the Z-boson mass, in other words near 
the Z-pole.
Due to the resonance production of Z-bosons the statistics is good
(several million events) and it allowed to measure electroweak (EW)
parameters with high precision. 

Any parity violating interactions eventually give rise to the space
asymmetryand could be revield
in, for example,  forward-backward asymmetry measurement. 
Axial-vector type interactions of torsion with matter fields is
this case of interactions.
But the  source of asymmetry also exists in the SM EW interactions  
because of the presense of the $\gamma_\mu\gamma_5$ structure in
the  interactions of $Z$- and $W$-bosons with  fermions.
The interactions between $Z$-boson and fermions can be written in
general form as:
\beq
L_{Zff}= -\frac{g}{2\, cos\theta_W}
\sum_i \,{\bar {\psi}}_i\,\gamma^\mu(g^i_V-g^i_A\gamma^5) \psi\, 
Z_\mu\,,
\label{zff}
\eeq
where, $\theta_W$ is Weinberg angle, $g=e/sin\theta_W$
 ($e$ - positron charge); 
and the vector and axial couplings are:
\begin{eqnarray}
g_V^i&\equiv& t_{3L}(i)- 2 q_i\, sin^2\theta_W,\\
g_A^i&\equiv& t_{3L}(i).
\label{ga}
\end{eqnarray}
Here $t_{3L}$ is the weak isospin of fermion
and $i$ has the values $+1/2$ for $u_i$ and $\nu_i$ while it is 
$-1/2$ for $d_i$ and $e_i$. Here
$i=1,2,3\,$ is the index of the fermion generation and
$q_i$ is the charge of the $\psi_i$ in  units of charge of positron.

The left handed fermion fields $\psi_i = $
\( \left( \begin{array}{c}  \nu_i\\ e^-_i  \end{array}\right) \)
 and 
\( \left(\begin{array}{c}   u_i\\ d'_i    \end{array}\right) \)
of the i$^{th}$ fermion family 
transform as a doublets under SU(2),
where ${d'}_i\equiv\sum_j V_{ij} d_j$ and  $V$ is the 
Cabibbo-Kobayashi-Maskawa mixing matrix.

The forward-backward asymmetry for $e^+e^- \rightarrow l^+l^-$ 
is defined as 
\begin{equation}
A_{FB}\equiv \frac{\sigma_F-\sigma_B}{\sigma_F+\sigma_B},
\label{asym}
\end{equation}
where $\sigma_F(\sigma_B)$ is the cross section for $l^-$
to travel forward(backward) with 
respect to electron direction. Such an asymmetries are measured
 at LEP~\ref{ewdata}.

{}From asymmetries one derives the ratio $g_V/g_A$ of vector
and axial-vector couplings.
Presence of torsion would  change  the forward-backward asymmetry and
would, as we show below, brightly reviel itself. 
In fact, the measured EW parameters are in a good agreement with the
theoretical predictions and hence one can establish the limits on the
torsion parameters based on the experimental errors.
The latest  relevant electroweak data are presented in 
Table~\ref{ewdata}.
\begin{table}[t]
\vspace{0.2cm}
\begin{center}
\begin{tabular}{|c|c|c|c|}
\hline
\hline
$A{FB}$& $A^l_{FB}$&   $A^b_{FB}$&   $A^c_{FB}$\\ 
   \hline
data   & 0.0171(10)&   0.0984(24)&    0.0741(48)\\
   \hline
   \hline
&&&\\
$g_{V}$& $g_{Ve}$&     $g_{V\mu}$&   $g_{V\tau}$\\
   \hline
data   &-0.0367(15) & -0.0374(36)&  -0.0367(15)\\
   \hline
    \hline
&&&\\
$g_{A}$& $g_{Ae}$&     $g_{A\mu}$&   $g_{A\tau}$ \\ 
   \hline
data&   -0.50123(44)& -0.50087(66)& -0.50102(74)\\
   \hline
   \hline
\end{tabular}
\end{center}
\label{ewdata}
\caption{\it
Results of combined LEP EW measurements for forward-backward asymmetry 
at Z-pole
and vector and axial couplings. 
$A^l_{BF}$ -- combined data for electron, muon and tau-lepton
asymmetries; $A^b_{BF}$ and  $A^c_{BF}$  -- 
asymmetries for b-quark and c-quark respectively.}
\end{table}

\begin{figure}[htb]
   \begin{center}
    \vskip -1cm\hspace*{-3cm}
    \epsfxsize=12cm\epsffile{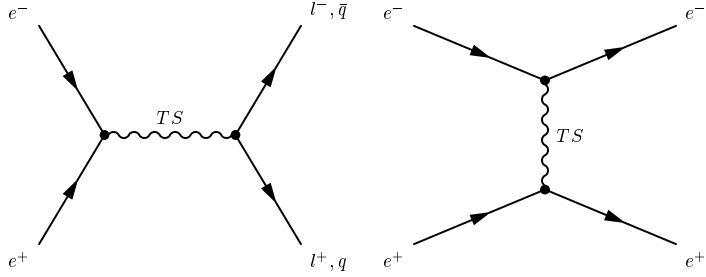}
    \vskip -12cm
\caption{\it Diagrams with torsion exchange in $e^+e^-$
collisions contributing to 
forward-backward lepton and quark asymmetry. TS indicates to the 
torsion propagators.}
    \vskip -0.5cm
\label{eediag}
\end{center}
\end{figure}

Hereafter we will use   $A^l_{BF}$ value for combined lepton asymmetry
and in particular for the 
electron $A^e_{BF}$ asymmetry in 
establishing the limits on the torsion parameters.
We have calculated the contribution to the asymmetries from
torsion exchange diagrams shown in Fig.~\ref{eediag}.  

\begin{figure}[htb]
  \vspace*{0.5cm}
    \epsfxsize=16cm
    \epsffile{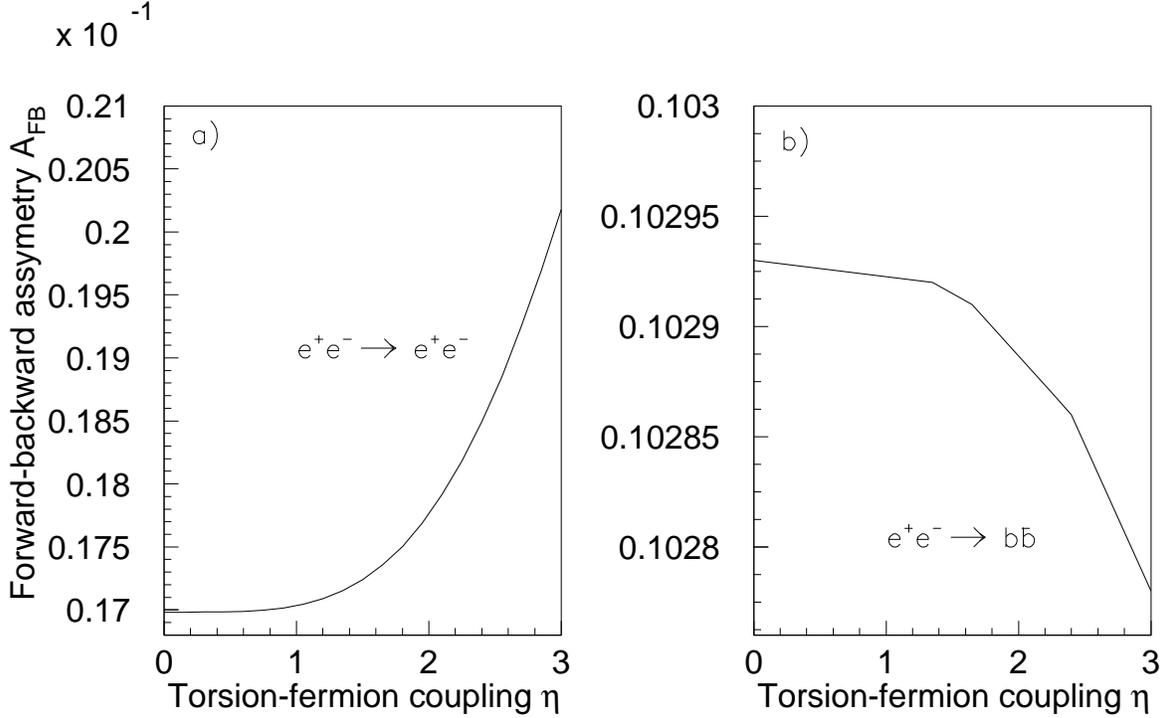}
 \vspace*{-1.0cm}
    \caption{\label{fig:lepasymm} {\it Forward-backward
$\,e^-e^+\rightarrow e^-e^+\,$ and $\,e^-e^+\rightarrow b\bar{b}\,$ 
asymmetries versus
    torsion $\,\eta\,$ coupling}}
 \end{figure}

{}From those calculations we have found limits at $95\%$ confidence
level (CL) on the
$\eta$ coupling for different torsion masses taking into account the 
error
of the experimental measurements.
The most precise individual measurement at LEP is  from $A^b_{FB}$.
But it turned out that  the most restrictive limit on $\eta$ comes from
the electron asymmetry of $e^-e^+\rightarrow e^-e^+$ scattering 
since
in this case both ($s$- and $t$-channel) diagrams from the 
Fig.~\ref{eediag}
contribute to $A^e_{FB}$. In the case of $e^-e^+\rightarrow b\bar{b}$
only $s$-channel diagram (the first one on the Fig.~\ref{eediag})
does contribute to the $A^b_{FB}$ asymmetry.
$A^e_{FB}$  and $A^b_{FB}$ $Z$-pole asymetries for $Z$-bozon
exchange diagram and for the diagrams with the torsion exchange, 
shown in Fig.~\ref{eediag} can be written as follows:
\begin{equation}
A_{FB}^{e(b)}=\frac{(\sigma_F-\sigma_B)^{e(b)}}{(\sigma_F+\sigma_B)^{e(b)}},
\end{equation}
where $(\sigma_F-\sigma_B)^{e(b)}$ and $(\sigma_F+\sigma_B)^{e(b)}$ are 
the
following expressions:
\begin{eqnarray}
( \sigma_F-\sigma_B)^e & = &K_l\left[ 
\frac{e^4 {g_a^e}^2 {g_v^e}^2 M_Z^2}{ 2 c_W^4 s_W^4 \Gamma_Z^2}+
\frac{e^2 \eta^2 M_{TS} }{ 2 c_W s_W\Gamma_Z}
[ ({g_a^e}^2+{g_v^e}^2) F_1^{FB} - 2{g_v^e}^2 F_2^{FB}] + \eta^4 
F_3^{FB}\right] \\
(\sigma_F-\sigma_B)^b & = &K_q\left[ 
\frac{e^4 {g_a^e}{g_a^b}{g_v^e}{g_v^b}M_Z^2}{ 2 c_W^4 s_W^4 \Gamma_Z^2}-
\frac{4 e^2 \eta^2 M_{TS} }{ 3 c_W s_W\Gamma_Z}
 {g_v^e} {g_v^b} F_2^{FB} \right] \\
(\sigma_F+\sigma_B)^e&=&K_l \left[ 
\frac{e^4 ({g_a^{e}}^2+{g_v^{e}}^2)^2 M_Z^2}{ 6 c_W^4 s_W^4 \Gamma_Z^2}+
\frac{e^2 \eta^2 M_{TS} }{ 2 c_W s_W\Gamma_Z}
[ ({g_a^e}^2+{g_v^e}^2) F_1^{T} - \frac{8}{3}{g_v^e}^2 F_2^{T}] + \eta^4 
F_3^{T} \right] \\
(\sigma_F+\sigma_B)^b&=&K_q\left[ 
\frac{e^4 ({g_a^{e}}^2+{{g_v^{e}}^2})({g_a^{b}}^2+{g_v^{b}}^2) M_Z^2}{ 6 
c_W^4 s_W^4 \Gamma_Z^2}-
\frac{4 e^2 \eta^2 M_{TS} }{ 3 c_W s_W\Gamma_Z}
{g_v^e}{g_v^b} F_2^{T}\right] , 
\end{eqnarray}
Here $s_W$ and $c_W$ are $sin$ and $cos$ of the Weinberg angle,
$\Gamma_Z$ -- width of $Z$-boson, $K_l=1/(16\pi M_Z^2)$, \ \ 
$K_q=3/(16\pi M_Z^2)$, \ \
$F_{1,2,3}^{FB(T)}$ function are written as follows:
\begin{eqnarray}
&&     F_1^{FB}= -z - \frac{4(1+z^2)^2 y_1}{z^3}, \nonumber \\ 
&&F_2^{FB}= F_2^T =\frac{z^3}{z^2-1}, \nonumber\\
&&F_3^{FB}=\frac{4 z^2(1+2z^2+2z^4)}{(z^2+1)(z^2+2)}
            +\frac{2 z^2}{z^2-1}  + \frac{8(3+z^2)}{z^2-1}y_1\\
&&     F_1^{T}= \frac{-z^2(2+3z^2)+2(1+z^2)^2 y_2}{z^3}, \nonumber\\
&&F_3^{T}=4(1+z^2) - \frac{2z^2}{1+z^2}  +
             \frac{8z^4}{3 (1-z^2)^2}+\frac{4(2+3z^2)-8(1+z^2) 
y_2}{(z^2-1)},
\end{eqnarray}
where
\begin{eqnarray}
&&z= \frac{M_Z}{M_{TS}}\,,\, \,\, \,\, 
y_1=ln\frac{1+z^2}{(1+z^2/2)^2}\,,\, \,\, \,\,  y_2=ln(1+z^2)\,.
\end{eqnarray}
When torsion exchange is absent ($\eta=0$), then from 
the formulaes below one obtains a
well known  result~\cite{ewdata} for tree level SM asymmetries:\\
\begin{eqnarray}
&&A^{b(e)}_{FB}=\frac{3{g_a^e}{g_a^{b(e)}}{g_v^e}{g_v^{b(e)}}}
                 {({g_a^{e}}^2+{g_v^{e}}^2)({g_a^{b(e)}}^2+{g_v^{b(e)}}^2)}
\end{eqnarray}
Electroweak radiative corrections can be absorbed into $g_a$ and $g_v$ 
values
formulae written above  will be valid including loop corrections 
for redefined  $g_a$ and $g_v$ values.
Figure~\ref{fig:lepasymm}a) and b) shows the behavior  of the 
$A^e_{FB}$ and  $A^b_{FB}$ asymmetries respectively versus 
$\eta$ torsion coupling when 
torsion mass is fixed to 1 TeV. 
We use  electroweak parameters of $M_Z$, $\Gamma_Z$, $sin\theta_W$,
$g_{A(V)}$ from \cite{ewdata}.
One can see that $A^b_{FB}$ asymmetry depends much weaker on $\eta$
and goes down with increasing of $\eta$ while $A^e_{FB}$ is  increasing 
with 
increasing of $\eta$.
For zero $\eta$
the asymmetries are equal to it's SM predictions
(which is slightly different 
from measured value, see \cite{ewdata} for details). They are
not zero because of the presence 
of the axial-vector coupling in the interactions of $Z$-boson and 
electron or quark.
Deviations of the asymmetry from SM
predictions would be indication of the  presense of the additional 
torsion-like type axial-vector  interactions. Our analysis shows that
$A^e_{FB}$ asymmetry is the best observable among others asymmetries to
look for torsion.

\begin{figure}[htb]
   \begin{center}
    \vskip -0.8cm\hspace*{-0.5cm}
    \epsfxsize=8cm\epsffile{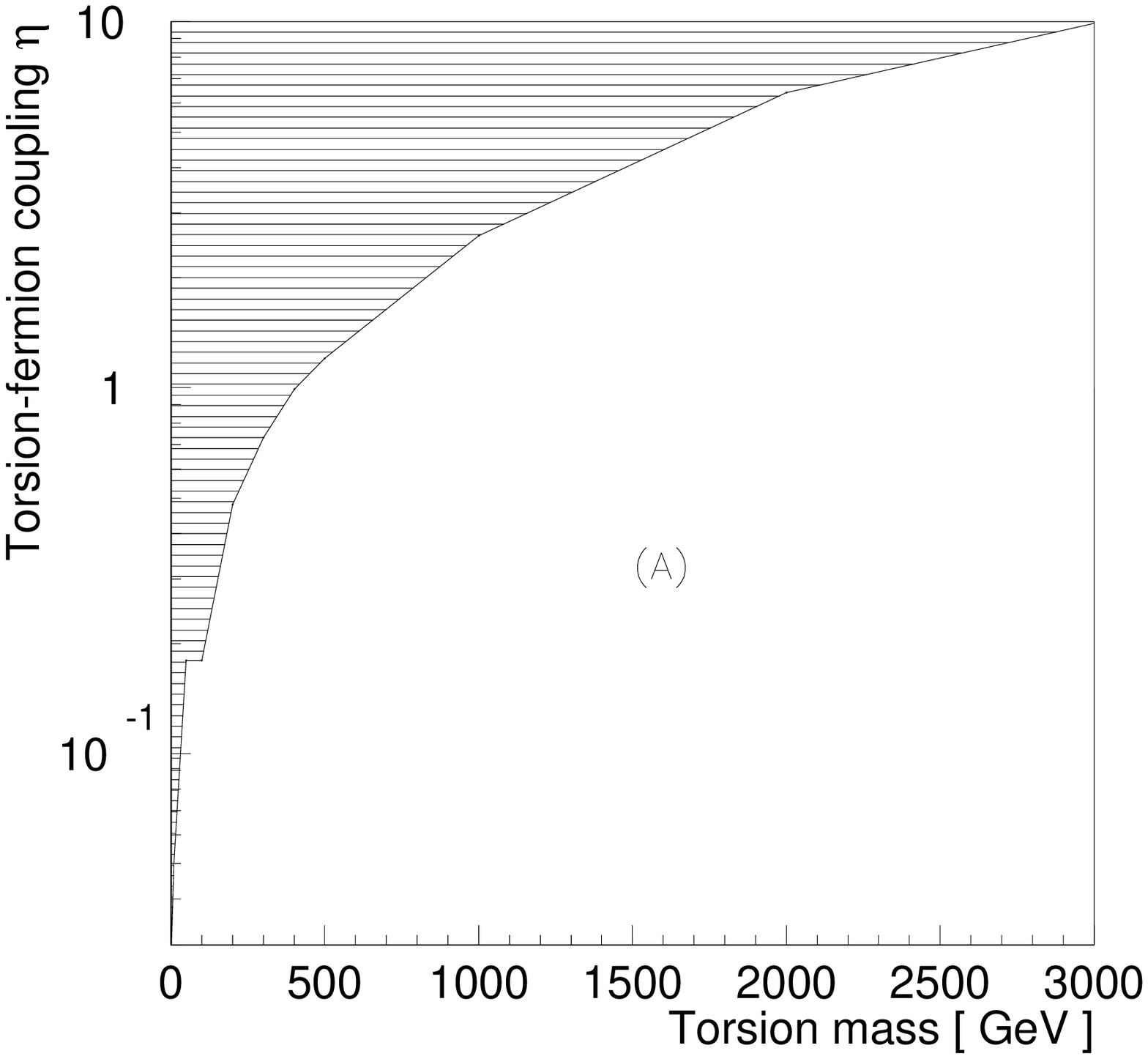}\epsfxsize=8cm\epsffile{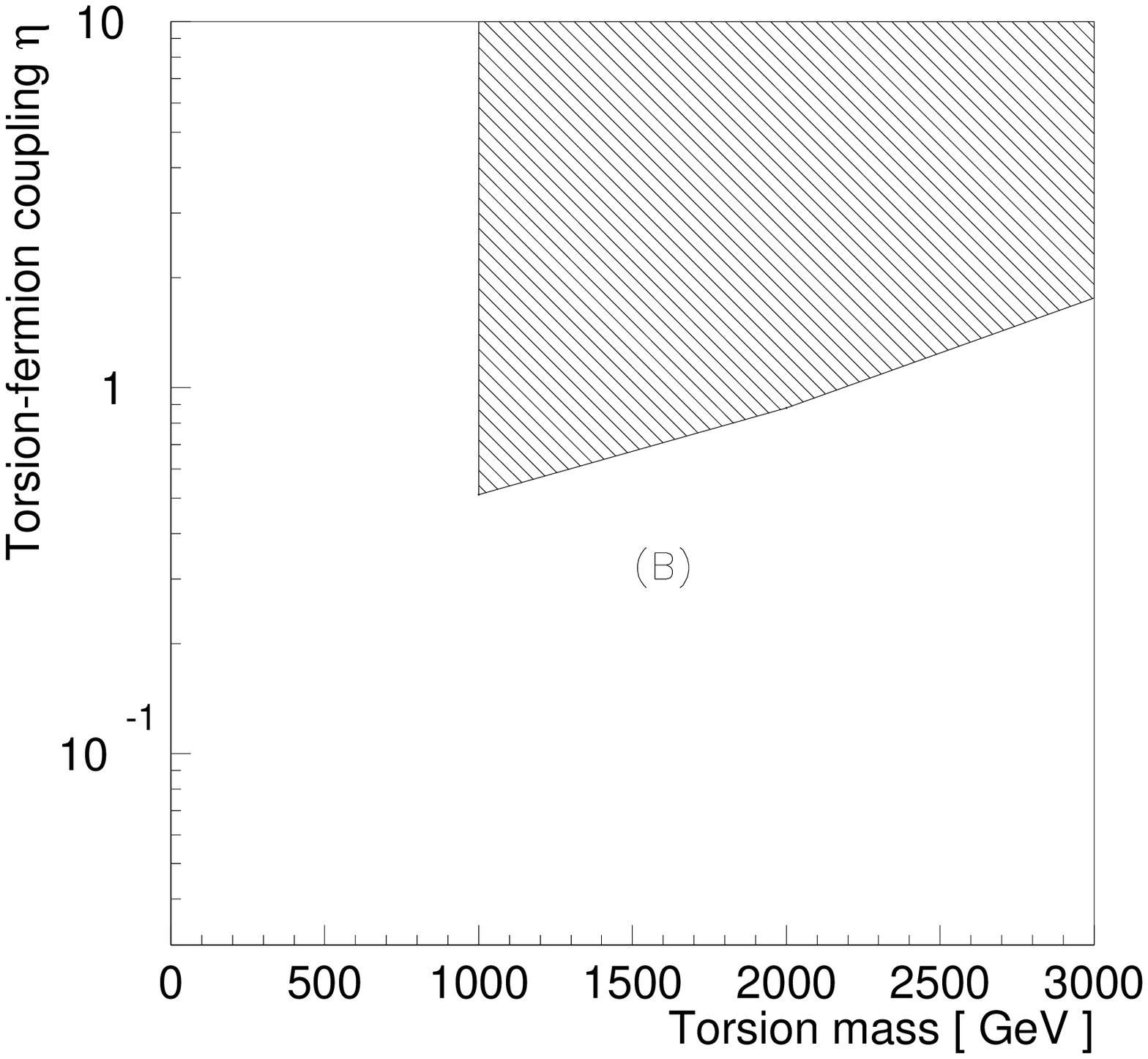}
    \vskip -0.8cm\hspace*{-0.5cm}
    \epsfxsize=8cm\epsffile{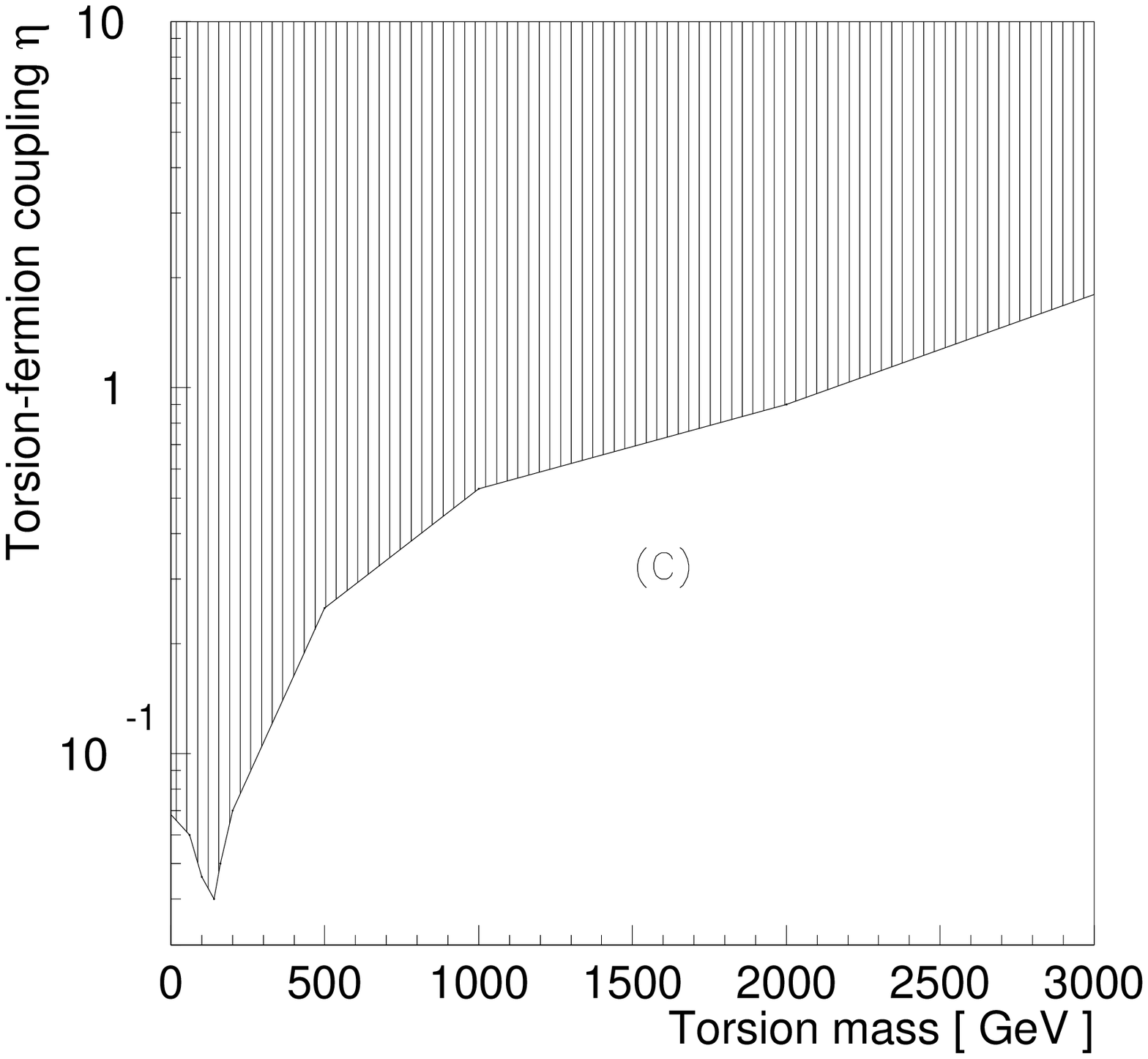}\epsfxsize=8cm\epsffile{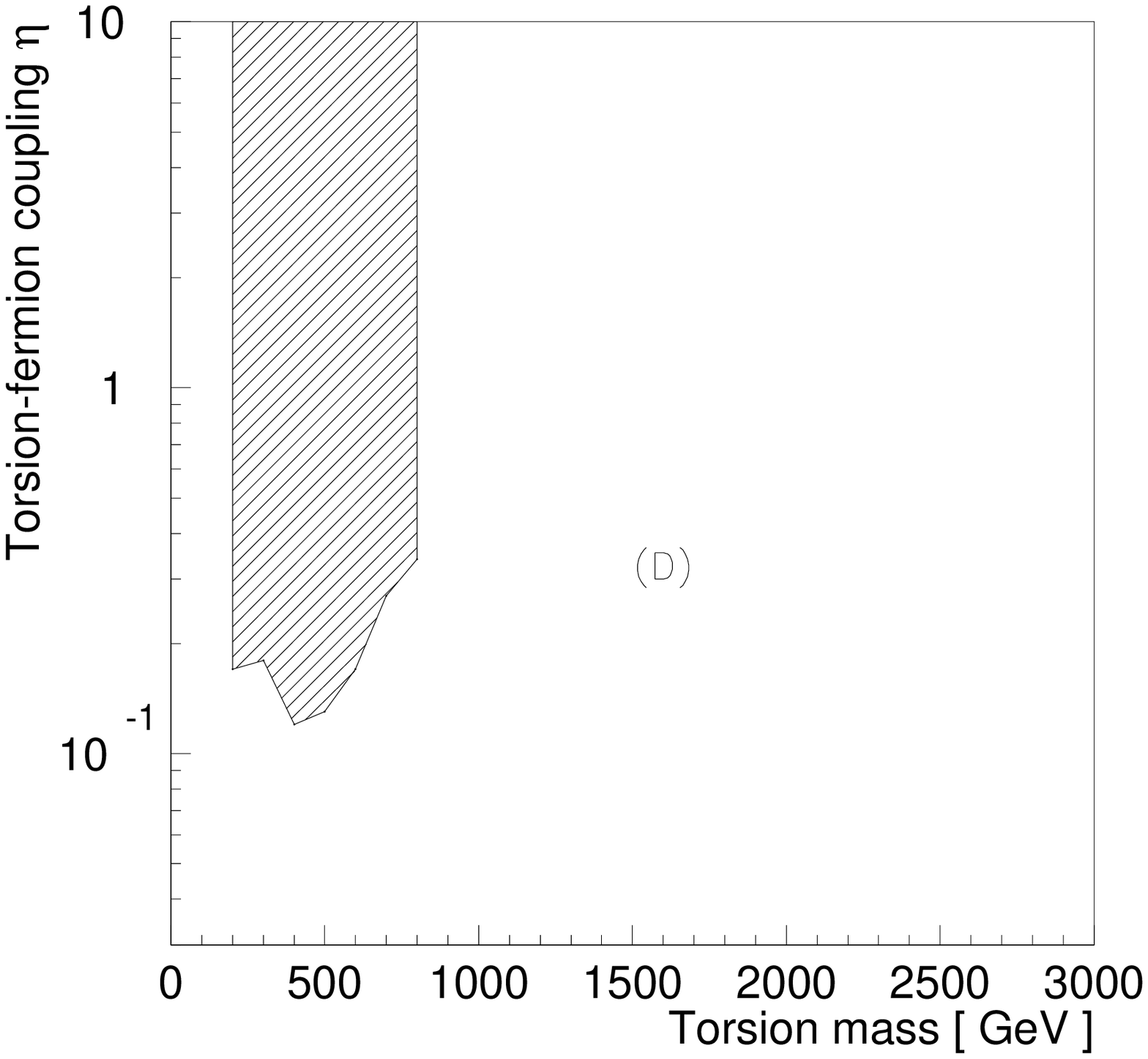}
    \vskip -0.8cm\hspace*{-0.5cm}
    \epsfxsize=8cm\epsffile{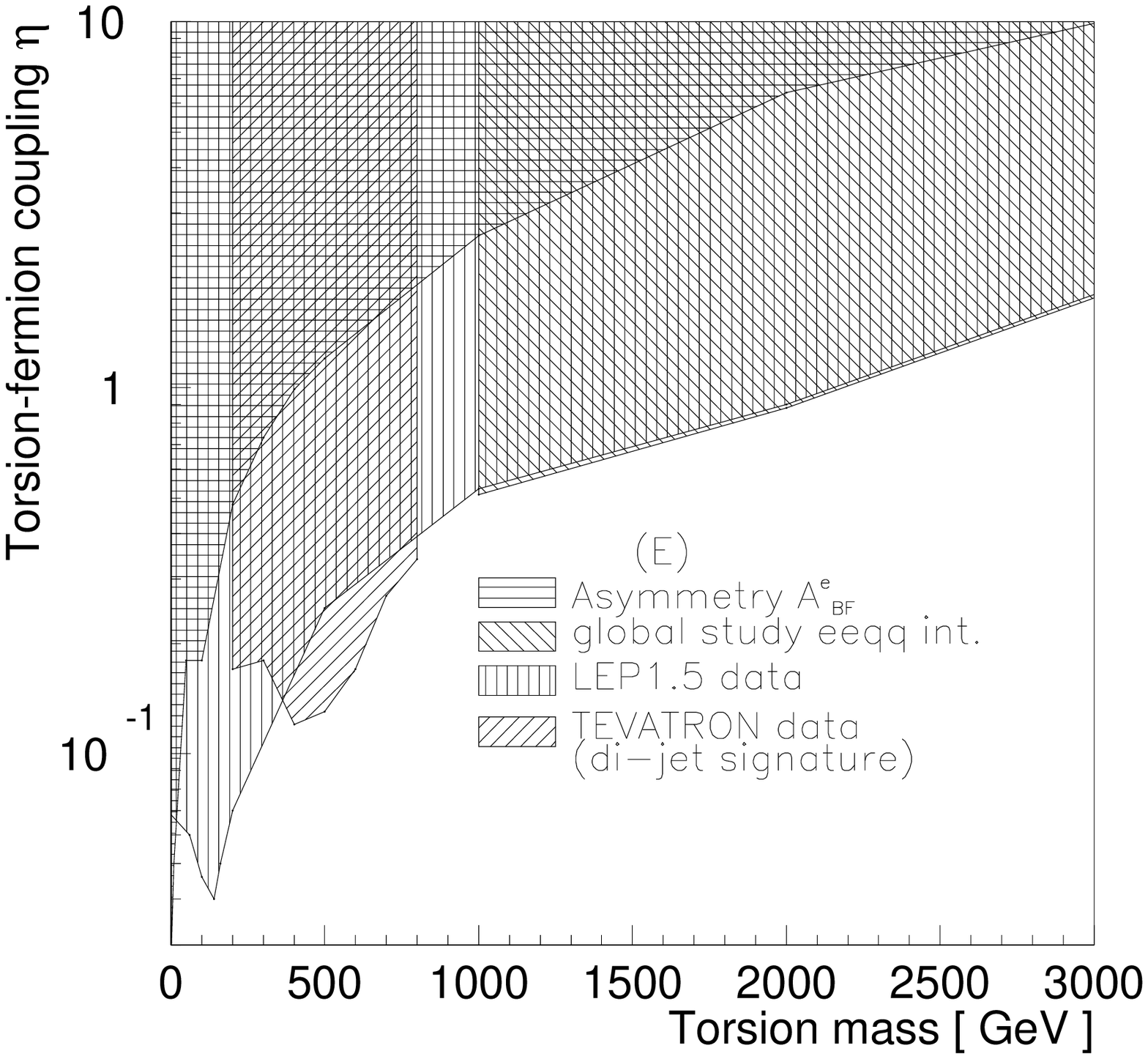}
    \\
  \end{center}
  \vspace*{-1.0cm}
    \caption{\it Allowed regions for $M_{ts}$ and  $\eta$ coming from
    LEP $A^e_{FB}$ asymmetry at Z-pole (A),
    global study of electron-quark contact interactions (B),
    LEP1.5 (C)  and TEVETRON data (D).
    (E) -- combined limit. Hatched region is excluded by   experiments
    mentioned above at 95\% CL}
    \label{fig:res}
\end{figure}

Exclusion region for $(\eta, M_{TS})$ plane coming from this asymmetry 
is
shown in Fig.~\ref{fig:res}(A).
Some  numbers corresponding to this limit are presented in
Table~\ref{tab:asym}.
\begin{table}[htb]
\begin{center}
\begin{tabular}{ | l | l | l | l | l | l | l | l |}
\hline
$M_{ts}$(GeV)& 1 & 10  &50  & 100  &200  & 1000 & 3000\\
\hline
$\eta$     &0.018&0.050&0.18&0.18  &0.48 &2.6   &9.9\\
\hline
\end{tabular}
\end{center}
\caption{\it Points for exclusion curve at 95\% CL  in ($M_{ts},\eta$) 
plane from LEP Z-pole $A_{FB}^e$ asymmetry. 
These numbers correspond to the Fig.~\ref{fig:res}(A).}
\label{tab:asym}
\end{table}

\vskip 6mm
\noindent
{\large\bf 8. Torsion and 4-fermion contact interactions}
\vskip 2mm

The straightforward consequence of the heavy torsion
interacting with fermion fields is the effective four-fermion contact
interaction of leptons and quarks (\ref{contact}).
Four-fermion interaction effectively appears
for the torsion with a mass much higher than the energy
scale available at present colliders. In this section we are going to
set an upper bounds for the single parameter ${M_{ts}}/{\eta}\,$
from (\ref{contact}) using modern data. There are several experiments
from which the constraints on the contact four-fermion interactions
come:

\vskip 1mm
\noindent
1)Experiments on polarized electron-nucleus scattering --
SLAC e-D scattering experiment\cite{slac}, Mainz e-Be scattering
experiment \cite{mainz} and bates e-C scattering experiment 
\cite{bates};
\vskip 0.8mm
\noindent
2)Atomic physics parity violations measures \cite{apv}
electron-quark coupling that are different from those tested at
high energy experiment
 provides alternative constraints on new physics.
\vskip 0.8mm
\noindent
3) $e^+e^-$ experiments  - SLD, LEP1, LEP1.5 and LEP2 (see for example
\cite{lep,opal,l3,aleph,lang});
\vskip 0.8mm
\noindent
4)Neutrino-Nucleon DIS experiments -- CCFR collaboration obtained model
independent constraint on the
effective $\nu\nu q q$ coupling \cite{ccfr}.
\vskip 1mm

Here  we consider a limits on the contact interactions induced by
torsion. The contact four-fermion interaction may be described by the
Lagrangian~\cite{eich} of the most general form:
\begin{equation}
L_{\psi'\psi'\psi\psi}=g^2\sum_{i,j=L,R}\sum_{q=u,d}\frac{\epsilon_{ij}}
{(\Lambda_{ij}^{\epsilon})^2}\,
(\bar \psi'_i\gamma_\mu \psi'_i)\,(\bar \psi_j\gamma^\mu \psi_j)
\label{cont}
\end{equation}
Subscripts i,j refer to different fermion helicities:
$\, \psi^{(')}_i= \psi^{(')}_{R,L}= (1\pm\gamma_5)/2\cdot \psi^{(')}\,$;
where $ \psi^{(')}$ could be quark or lepton; $\,\,\Lambda_{ij}\,$
represents the mass scale of the exchanged new particle; coupling 
strength
is fixed by the relation: $\,g^2/4\pi=1$,
the sign factor  $\,\epsilon_{ij}=\pm 1\,$
allows for either constructive or destructive interference with
the SM $\gamma$ and $Z$-boson exchange amplitudes.
The formula (\ref{cont}) can be successfully used for the study of the
torsion-induced contact interactions because it includes an axial-axial
current interactions as a particular case.

Recently the global study of the
electron-electron-quark-quark($eeqq$) interaction sector of the
SM~\cite{global} have been done using data from all mentioned 
experiments.
The limits established in this paper
are the best in comparison with the previous ones.

The specific distinguishing feature of the contact interactions
induced by torsion
is that those contact interactions are of axial-axial type.
Therefore we used the limits obtained in paper~\cite{global}
for this kind of interaction. Limits on axial-axial ($AA$) type contact
interactions
mainly come from OPAL collaboration. It was shown (see for details
\cite{opal})
that present LEP data are particulary sensitive to $VV$ and $AA$ models
which could be distinguished from others types of contact interactions 
by analysing  of scattering angle distributions of outgoing leptons
and quarks.
The other possibility of study the chiral structure of contact 
interactions 
through  the polarized lepton and proton beams scattering analysis
can be realized  at HERA~\cite{hera}.

Axial-axial current may be expressed through
$\,LL, RR, LR\,\, {\rm and}\,\,  RL\,$ currents in the
following way:
 \begin{equation}
j_\mu^A j_\mu^A=
\frac{j_\mu^L j_\mu^L+j_\mu^R j_\mu^R-j_\mu^L j_\mu^R-j_\mu^R 
j_\mu^L}{4}\,.
\label{aa}
\end{equation}
For the axial-axial
$\,eeqq\,$ interactions~(\ref{cont}) takes the form (we put $g^2=4\pi$) 
:
\beq
L_{eeqq}=-\frac{4\pi}{{(\Lambda_{AA}^{\epsilon})^2}}
(\bar e\gamma_\mu\gamma_5 e)(\bar q\gamma^\mu \gamma_5 q)
\label{contaa}
\eeq
The limit for the contact axial-axial $eeqq$ interactions comes
from the global analysis of Ref.~\cite{global}:
\beq
\frac{4\pi}{\Lambda_{AA}^2} < 0.36\mbox{ TeV} ^{-2}
\label{glob}
\eeq
Comparing the parameters of the effective contact four-fermion
interactions of general form
(\ref{contaa}) and contact four fermion interactions induced by
torsion (\ref{contact}) we arrive at the following relations:
\beq
\frac{\eta^2}{M_{ts}^2}=\frac{4\pi}{{\Lambda_{AA}}^2}
\label{rel}
\eeq
{}From (\ref{glob}) and (\ref{rel}) one gets the following limit
on torsion parameters:
\beq
\frac{\eta}{M_{ts}}<0.6\mbox{ TeV}^{-1}\; \Rightarrow \;
M_{ts}>1.7\mbox{ TeV }\cdot\eta
\label{globfin}
\eeq
The limits on $M_{ts}$ and $\eta$ coming from the (\ref{globfin})
is shown in Figure~1(B). As we have already mentioned above the
restrictions concern only the ratio between the torsion mass and
coupling parameter. Some remark about the energy limits taken in this
plot is in order.
We started exclusion region from $M_{ts}=$~1~TeV. This choice is
related with the fact that the
application of effective-contact interactions (\ref{contact})
is valid up to the
certain mass of the torsion below which an exact calculation
(regarding the field $S_\mu$ as dynamical) should be done.
The relative data of the two approaches are shown on the Figure 2,
where the results for gauge interaction~(\ref{action})
and contact interactions~(\ref{cont}) for torsion are compared.
As an example  we have calculated total cross section for LEP1.5
with $\sqrt{s}=140$ GeV and $\eta$ equal to 0.5.
One can clearly see that for torsion heavier than 1 TeV the 
approximation
of the effective contact interaction works almost perfectly, reproducing
the result for the
exact calculation with 0.1\% accuracy. Therefore the scale 1 TeV
is appropriate starting point for  putting the limit on
torsion parameters using the Lagrangian with contact interactions.
\begin{figure}[htb]
  \vspace*{-1.0cm}
    \epsfxsize=12cm
    \epsffile{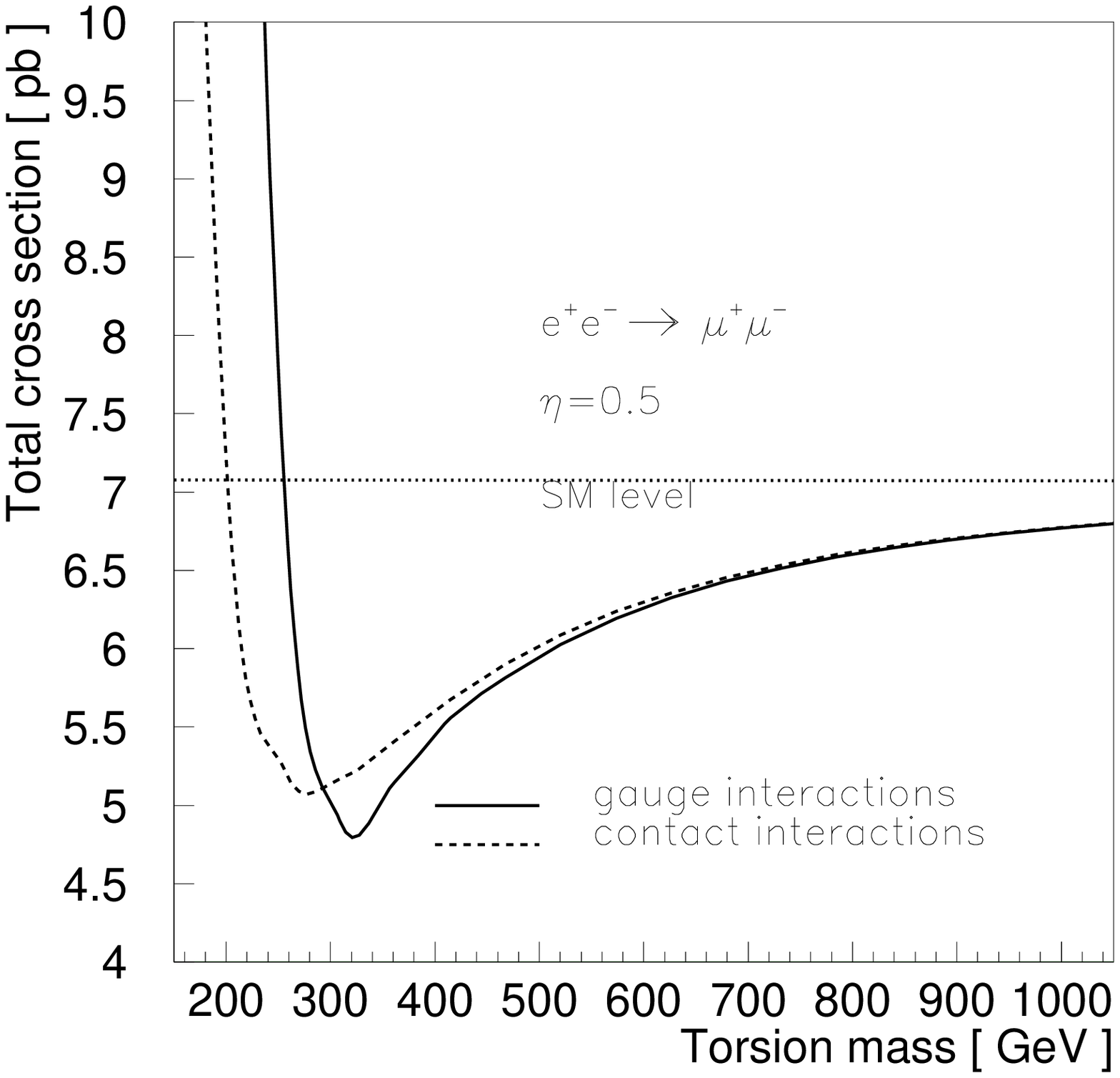}
 \vspace*{-1.0cm}
    \caption{\it Comparison  of the total cross sections
    of $e^+e^-\rightarrow \mu^+\mu^-$ process
    for gauge and contact interactions }
 \end{figure}
Scenario with light torsion is in general more difficult because here
we have two independent parameters and thus are enforced to
study the $2$-dimensional restrictions from the experimental data.
Indeed there is no rigid border between two cases, as we shall see 
below.

To consider the limits in the ($M_{ts},\eta$) parameter space for
the light torsion we use results of  LEP1.5 analysis of paper
\cite{opal}:
the cross section of $e^+e^-\rightarrow e^+e^-(\mu^+\mu^-)$ process
was measured with
accuracy 1-2\%. We used this fact  to put the limits on torsion mass
and coupling:
90\% acceptance for electron and 60\% for muon channels was
assumed, the total
cross section for these reaction
were calculated and 4\% deviation from the Standard
Model prediction was taken for establishing the limits.
The resulting constraints is shown in Figure~4(C).

\vskip 6mm
\noindent
{\large\bf 9. Limits on the torsion parameters from TEVATRON data}
\vskip 2mm

 The torsion with the mass
in the range of  present colliders could be produced in fermion-fermion
interactions as a resonance, decaying to fermion pair.
The most promising collider for search the signature
of such type is TEVATRON. This  proton-antiproton 1.8 TeV collider has 
the highes available
center of mass energy up to the moment.
So one can naturally expect that
the haviest resonance which could be produced in the quark-quark, 
quark-gluon 
or gluon-gluon collision would be discovered there.
 
 Search for New Particles Decaying to two-jets
has been done recently by D0 and CDF collaborations \cite{d0}.
The data that we use in our analysis are extracted from the
figure presented by the 
D0 collaboration which established the limit on the production
cross section of $Z'$ and $W'$ bosons.
Here we assume also 90\% events efficiency
(including efficiency of kinematical cuts
and trigger efficiency) and calculated the cross
section for torsion production at TEVATRON. 
For simplicity we also assumed that torsion coupling with only one 
kind of quark (u-quark) is nonzero and calculated the cross section of
the reaction $p\bar{p}(u\bar{u})\rightarrow TS\rightarrow u\bar{u}$.
Then we applied
D0 limit at 95\% CL
 for torsion production cross section and converted into  the
limit for  ($M_{ts},\eta$) plane. This limit is  shown in Fig.~4(D).
The points for the exclusion curve are given in Table~3.

\begin{table}[htb]
\begin{center}
\begin{tabular}{ |l | l | l | l | l | l | l | l| }
\hline
$M_{ts}$(GeV)& 200&300 &400 & 500 &600  & 700 & 800\\
\hline
$\eta$     &0.17&0.18&0.12&0.13&0.17&0.27&0.34\\
\hline
\end{tabular}
\end{center}
\caption{\it Points for exclusion curve in ($M_{ts},\eta$) plane
from TEVATRON data( 95\% CL). See Fig.~4(D).}
\end{table}
One can see that the limits on $\eta$ coming from these analysis
are better in comparison with those from the LEP data for some 
values of $\,\eta\,$ parameter.
Combined exclusion plot for $\,$($M_{ts},\eta$)$\,$ 
plane is presented in Fig.~4(E).

\vskip 6mm
\noindent
{\large\bf 10. Conclusions}
\vskip 2mm

Starting from the fermion-torsion coupling we have
derived the action of the propagating torsion and implemented
it into the abelian sector of the Standard Model. It was shown that the
only one action of torsion which leads to consistent effective field
theory includes propagating pseudovector massive particle with softly
broken (new) gauge symmetry. The renormalization group gives a strong
argument against the light torsion, because, due to the universality
of torsion-fermion interaction, light torsion means an unnaturally fast
running of the torsion mass. Since the scalar fields and Yukawa
coupling are inconsistent with propagating torsion, we base the
phenomenological part of our work on the fermion sector of the
SM only. In this way, we have established some upper
bounds on the torsion mass and torsion-fermion coupling constant
(which is supposed to be universal) using combined limit for
asymmetry of the forward-backward scattering, four-fermion contact
interactions and also LEP and TEVATRON data.
For  heavy torsion the limit is described by
relation~(\ref{globfin}) while for light torsion with the mass below 1 
TeV
limits coming from LEP and TEVATRON data  bound $\eta$ to be less than
0.1-0.02 depending on $M_{ts}$.

 Results presented above clearly show, 
that the best limits for light torsion could be obtained from LEP
asymmetry data. In particular for the $M_{ts} <$ 1 GeV 
the value of $\eta$ is less than 0.02. This gives an essential 
improvement 
as compared to our previous report \cite{betor} where this kind of 
observables was not considered.
The limits of the same order
or may be even better could be  obtained from data on the 
measurement of
the anomalous magnetic moment for electron and muon~\cite{kalm}.
At the same time the best limits for heavy torsion
come from global analysis of contact interactions~\cite{global}.

In the case of a very heavy mass $\,M_{ts} \sim M_{Pl}\,$
the torsion manifests itself only as an extremely weak contact 
interaction. The propagation and quantum effects of such 
a torsion may be described only in the framework of string theory.
In other words such a 
``very heavy'' torsion doesn't exist as an independent 
field. In this paper we have studied the alternative option.
 
\vskip 5mm
\noindent
{\bf Acknowledgments}
\vskip 2mm

Authors are grateful
to M. Asorey, I.L. Buchbinder, J.A. Helayel-Neto,
I.B. Khriplovich and T. Kinoshita for
stimulating discussions. We are also indebted to M. Kalmykov
for sharing with us the results of calculations
\cite{kalm} prior to publication.
\ I.\ L.\ Sh. acknowledges warm hospitality of Departamento de 
F\'{\i}sica,
Universidade Federal de Juiz de Fora and partial support by CNPq 
(Brazil)
and by Russian Foundation for Basic Research under the
project No.96-02-16017. A.\ S.\ B.\ is grateful to the Instituto de
F\'{\i}sica Te\'orica for its kind hospitality and acknowledges support
from Funda\c{c}\~ao de Amparo \`a Pesquisa do Estado de S\~ao Paulo
(FAPESP).


\newpage
\begin {thebibliography}{99}

\bibitem{GSW} M.B. Green, J.H. Schwarz  and E. Witten,
{\it Superstring Theory} (Cambridge University Press, Cambridge, 1987).

\bibitem{dat} B.K. Datta,{\sl Nuovo Cim.} {\bf 6B} (1971) 1; 16.

\bibitem{hehl} F.W. Hehl,
Gen. Relat.Grav.{\bf 4}(1973)333;{\bf5}(1974)491;

F.W. Hehl, P. Heide, G.D. Kerlick and J.M. Nester,
      Rev. Mod. Phys.{\bf 48} (1976) 3641.

\bibitem{aud} J. Audretsch, {\sl Phys.Rev.} {\bf 24D} (1981) 1470.

\bibitem{prec} H. Rumpf,{\sl Gen. Relat. Grav.} {\bf 14} (1982) 773.

\bibitem{kernel} {Goldthorpe W.H.,{\sl Nucl. Phys.}{\bf 170B} (1980) 
263};

{Cognola G., Zerbini S.,{\sl Phys.Lett.}{\bf 214B} (1988) 70};

{Gusynin V.P.,{\sl Phys.Lett.}{\bf 225B} (1989) 233};

{Nieh H.T., Yan M.L.,{\sl Ann. Phys.}{\bf 138} (1982)237}.

\bibitem{creation} H. Rumpf, {\sl Gen.Rel.Grav.} {\bf 10}
(1979) 509; 525; 647.

\bibitem{hehl-review}
"On the gauge aspects of gravity",
F. Gronwald, F. W. Hehl, GRQC-9602013, Talk given at International
School of Cosmology and Gravitation: 14th Course: Quantum Gravity,
Erice, Italy, 11-19 May 1995, gr-qc/9602013

\bibitem{bush1} I.L. Buchbinder and I.L. Shapiro,
{\sl Phys.Lett.} {\bf 151B} (1985)  263.

\bibitem{bush2} I.L. Buchbinder and I.L. Shapiro,
{\sl Class. Quantum Grav.} {\bf 7} (1990) 1197;

I.L. Shapiro, {\sl Mod.Phys.Lett.}{\bf 9A} (1994) 729.

\bibitem{babush} V.G. Bagrov, I.L. Buchbinder and I.L. Shapiro,
{\sl Izv. VUZov, Fisica (in Russian, English translation: Sov.J.Phys.)}
{\bf 35,n3} (1992) 5; see also hep-th/9406122.

\bibitem{hammond} R. Hammond, {\sl Phys.Lett.} {\bf 184A} (1994) 409;
{\sl Phys.Rev.} {\bf 52D} (1995) 6918.

\bibitem{doma1} A. Dobado and A. Maroto, 
{\sl Mod.Phys.Lett.} {\bf A12} (1997) 3003.

\bibitem{lamme} C. Lammerzahl, {\sl Phys.Lett.} {\bf 228A} (1997) 223.

\bibitem{rytor} L.H. Ryder and I.L. Shapiro, {\sl Phys.Lett.A}, 
to be published.

\bibitem{book} I.L. Buchbinder, S.D. Odintsov and I.L. Shapiro,
{\sl Effective Action in Quantum Gravity.} (IOP Publishing -- Bristol,
 1992).

\bibitem{rano} B. Nodland and J. Ralston, {\sl Phys.Rev.Lett.},
{\bf 78} (1997) 3043.

\bibitem{kibble} T.W. Kibble, {\sl J.Math.Phys.} {\bf 2} (1961) 212

\bibitem{nev} D.E. Nevill, {\sl Phys.Rev.} {\bf D18} (1978) 3535.

\bibitem{sene} E. Sezgin and P. van Nieuwenhuizen,
{\sl Phys.Rev.} {\bf D21} (1980) 3269.

\bibitem{zwei} Zwiebach B., {\sl Phys.Lett.} {\bf 156B} (1985) 315;

Deser S. and Redlich A.N., {\sl Phys.Lett.}{\bf 176B} (1986) 350;

Jones D.R.T., Lowrence A.M., {\sl Z.Phys.} {\bf 42C} (1989) 153.

\bibitem{weinberg} S. Weinberg, {\sl The Quantum Theory of Fields:
Foundations.} (Cambridge Univ. Press, 1995).

\bibitem{dogoho} J.F. Donoghue, E. Golowich and B.R. Holstein, 
{\sl Dynamics of the Standard Model} (Cambridge University Press, 1992).

\bibitem{volatyu} B.L. Voronov, P.M. Lavrov and I.V. Tyutin,
Sov.J.Nucl.Phys. {\bf 36} (1982) 498;

J. Gomis and S. Weinberg, Nucl.Phys. {\bf B469} (1996) 473.

\bibitem{don}Donoghue J.F., {\sl Phys.Rev.Lett.} {\bf 72} (1994) 2996;
{\sl Phys.Rev.}{\bf D50} (1994) 3874.

\bibitem{betor} A.S. Belyaev and I.L. Shapiro, {\sl Phys.Lett.} {\bf B},
to be published.

\bibitem{carroll} S.M. Caroll and G.B. Field,
{\sl Phys.Rev.} {\bf 50D} (1994) 3867.

\bibitem{kalm} M. Yu. Kalmykov, private communication.

\bibitem{novello} M. Novello, {\sl Phys.Lett.} {\bf 59A} (1976) 105.

\bibitem{apecor} T. Appelquist and J. Corrazone,
{\sl Phys.Rev.} {\bf D11} (1975) 2856.

\bibitem{vector} L.D. Faddeev and A.A. Slavnov,
{\sl Gauge fields. Introduction to quantum theory.}
(Benjamin/Cummings, 1980).

\bibitem{wein} S. Weinberg, {\sl Phys.Rev.} {\bf 118} 838.

\bibitem{votyu} B.L. Voronov and I.V. Tyutin, Sov.J.Nucl.Phys.
{\bf 23} (1976) 664.

\bibitem{buodsh} I.L. Buchbinder, S.D. Odintsov and I.L. Shapiro,
{\sl Phys.Lett.} {\bf 162B} (1985) 92.

\bibitem{hela} J.A. Helayel-Neto, {\sl Il.Nuovo Cim.} {\bf 81A} 
(1984) 533; J.A. Helayel-Neto, I.G. Koh and H. Nishino,
{\sl Phys.Lett.} {\bf 131B} (1984) 75.

\bibitem{stelle}  K.S. Stelle, Phys.Rev. {\bf 16D}, 953 (1977).

\bibitem{unit} B.\ Lee, C.\ Quigg and H.\ Thacker, Phys.\ Rev.\
Lett.\ {\bf 38}, 883 (1977); Phys.\ Rev.\ {\bf D16}, 1519 (1977);
D.\ Dicus and V.\ Mathur, Phys.\ Rev.\ {\bf D7}, 3111 (1973).

\bibitem{ewdata}
{\it The LEP Electroweak Working Group}, CERN-PPE/97-154 (1997).

\bibitem{comp}
     	E.E.Boos, M.N.Dubinin, V.A.Ilyin, A.E.Pukhov, V.I.Savrin,
	SNUTP-94-116, INP-MSU-94-36/358,
	{\tt hep-ph/9503280};  \\
    	P.A.Baikov {\it et al.}, Proc. of X Workshop on HEP and QFT
	(QFTHEP-95), ed. by B.Levtchenko, V.Savrin, p.101,
	{\tt hep-ph/9701412}.

\bibitem{slac} C.Y. Prescott et al., Phys. Lett. {\bf B84}, 524 (1979)

\bibitem{mainz} W. Heil  et al., Nucl. Phys. {\bf B327}, 1 (1989)

\bibitem{bates} WP.A. Souder  et al., Phys.Rev.Lett. {\bf 65}, 694 
(1990)

\bibitem{apv}
P.Langasker, M.Luo and A.Mann, Rev.Mod.Phys. {\bf 64}, 86 (1992)

\bibitem{lep}LEP Collaborations and SLD Collaboration, ``A Combination 
of
Preliminary Electroweak Measurements and Constrains on
the Standard Model'',
prepared from contributions to the 28th International Conference on High
Energy Physics, Warsaw,  Poland, CERN-PPE/96-183 (Dec. 1996).

\bibitem{opal}
OPAL Collaboration, G.~Alexander et al., {\bf B391}, 221 (1996).

\bibitem{l3}
L3 Collaboration, Phys. Lett. {\bf B370}, 195 (1996);
CERN-PPE/97-52, L3 preprint 117 (May 1997).

\bibitem{aleph}
ALEPH Collaboration, Phys. Lett. {\bf B378}, 373 (1996).

\bibitem{lang}
P. Langacker and J. Erler, presented at the Ringberg Workshop on the
Higgs Puzzle, Ringberg, Germany, 12/96, hep-ph/9703428.

\bibitem{ccfr}K.S. McFarland et al. (CCFR), FNAL-Pub-97/001-E,
hep-ex/9701010.

\bibitem{hera}
J.M. Virey, CPT-97-P-3542, To be published in
the proceedings of 2nd Topical Workshop on Deep Inelastic Scattering off
Polarized Targets: Theory Meets Experiment (SPIN 97), Zeuthen,
Germany, 1-5 Sep 1997: Working Group on 'Physics with
"Polarized Protons at HERA" ,hep-ph/9710423

\bibitem{global}
V. Barger, K. Cheung, K. Hagiwara, D. Zeppenfeld;
MADPH-97-999, hep-ph/9707412.
{\sl We are thankful to the authors of this paper
for their analysis}.

\bibitem{eich} E.Eichten, K.Lane, and M.Peskin, Phys.Rev.Lett.
{\bf 50}, 811
(1983).

\bibitem{d0}
 CDF Collaboration and D0 Collaboration (Tommaso Dorigo
for the collaboration),
FERMILAB-CONF-97-281-E, 12th Workshop on Hadron Collider
Physics (HCP 97), Stony Brook, NY, 5-11 Jun 1997

\end{thebibliography}
\end{document}